\documentclass[aps,prl,reprint,
 amsmath,amssymb,longbibliography
]{revtex4-2}
\usepackage{graphicx}
\usepackage{textcomp}
\usepackage{amssymb}
\usepackage[utf8]{inputenc}
\usepackage{amsmath}
\usepackage{bm}
\usepackage{tipa}
\usepackage{comment}
\usepackage{mathtools}
\usepackage{braket}
\usepackage{listings}
\usepackage{xcolor}
\usepackage{soul}
\usepackage{etoolbox}
\usepackage{lineno}
\usepackage[normalem]{ulem}
\usepackage[colorlinks,linkcolor=blue,anchorcolor=blue,citecolor=blue,urlcolor=black]{hyperref}

\definecolor{Darkgreen}{rgb}{0,0.4,0}
\definecolor{listinggray}{gray}{0.9}
\definecolor{lbcolor}{rgb}{0.9,0.9,0.9}
\begin{document}

\title{Observation of strong electric dipole-dipole interactions between two atoms}

\author{Yu-Kun Lu}
\thanks{Corresponding author. E-mail: yukunlu@mit.edu.}
\altaffiliation[Present address: ]{Department of Physics, Harvard University, Cambridge, Massachusetts 02138, USA}
\affiliation{Department of Physics, Massachusetts Institute of Technology, Cambridge, MA 02139, USA}
\affiliation{Research Laboratory of Electronics, Massachusetts Institute of Technology, Cambridge, MA 02139, USA}
\affiliation{MIT-Harvard Center for Ultracold Atoms, Cambridge, MA, USA}

\author{Jiahao Lyu}
\affiliation{Department of Physics, Massachusetts Institute of Technology, Cambridge, MA 02139, USA}
\affiliation{Research Laboratory of Electronics, Massachusetts Institute of Technology, Cambridge, MA 02139, USA}
\affiliation{MIT-Harvard Center for Ultracold Atoms, Cambridge, MA, USA}

\author{Wolfgang Ketterle}
\affiliation{Department of Physics, Massachusetts Institute of Technology, Cambridge, MA 02139, USA}
\affiliation{Research Laboratory of Electronics, Massachusetts Institute of Technology, Cambridge, MA 02139, USA}
\affiliation{MIT-Harvard Center for Ultracold Atoms, Cambridge, MA, USA}

\begin{abstract}
We have prepared pairs of atoms at densities exceeding $  10^{16}/\rm{cm}^3$ corresponding to $n\lambdabar^3>10$ and studied how they interact with light. In this regime, the coherent part of the electric dipole-dipole interaction is larger than the dissipative part, and a single photon emission can transfer large momenta (larger than ten times the single photon recoil) to the atoms. For off-resonant laser light, the atoms are dressed with an electric dipole moment of $\approx$ 1 Debye and interact coherently. This light-induced dipole-dipole interaction has been previously observed only in the far field mean-field regime. For two atoms, the resonance is shifted by the two-body Lamb shift which is dominated by the near-field part of the dipolar interaction. Dipolar resonance shifts have been observed before only for many atoms interacting with the far-field part of the dipolar interaction. Our work sheds new light on photoassociation processes and paves the way for studying driven quantum systems with strong electric dipole-dipole interactions. 
\end{abstract}
\maketitle

For a single atom, coupling to the electromagnetic vacuum results in energy-level shifts (the Lamb shift) and spontaneous emission of excited states~\cite{API}. It serves as a paradigmatic system for studying atom-light interaction, and the precise measurement of the Lamb shift triggered the development of quantum electrodynamics~\cite{PhysRev.72.241}. 
For many atoms, their coupling to a shared electromagnetic environment leads to electric dipole-dipole interactions between them~\cite{lehmberg1970radiation}. This electric dipole-dipole interaction is strong in the near-field regime ($kr\ll1$), but has been mainly studied in the far-field regime ($kr\gg1$) where the weak interactions between many atoms add up to an observable effect which is often classical. This includes the observation of deformation of expanding Bose-Einstein condensates (BEC) caused by light-induced dipole-dipole interactions~\cite{maiwoger2022observation}, the spectroscopic shift (collective Lamb shift) and broadening of the resonance for cold atoms in dense clouds~\cite{PhysRevLett.113.133602,PhysRevLett.116.233601,PhysRevLett.116.183601,bromley2016collective,PhysRevLett.117.073003,PhysRevA.94.023842,PhysRevA.97.053816} and regular arrays~\cite{PhysRevLett.113.193002,PhysRevLett.124.253602,rui_sub_radiant_mirror_2020,hutson2024observation,xf8r-ymrj}. Near-field studies in the optical domain were limited to experiments using inhomogeneously broadened dense hot vapors~\cite{weller2011absolute,PhysRevLett.108.173601,PhysRevLett.120.243401}, and atom loss through photoassociation induced by a far detuned laser beam~\cite{RevModPhys.78.483}.

\begin{figure}
    \centering
\includegraphics[width=0.8\columnwidth]{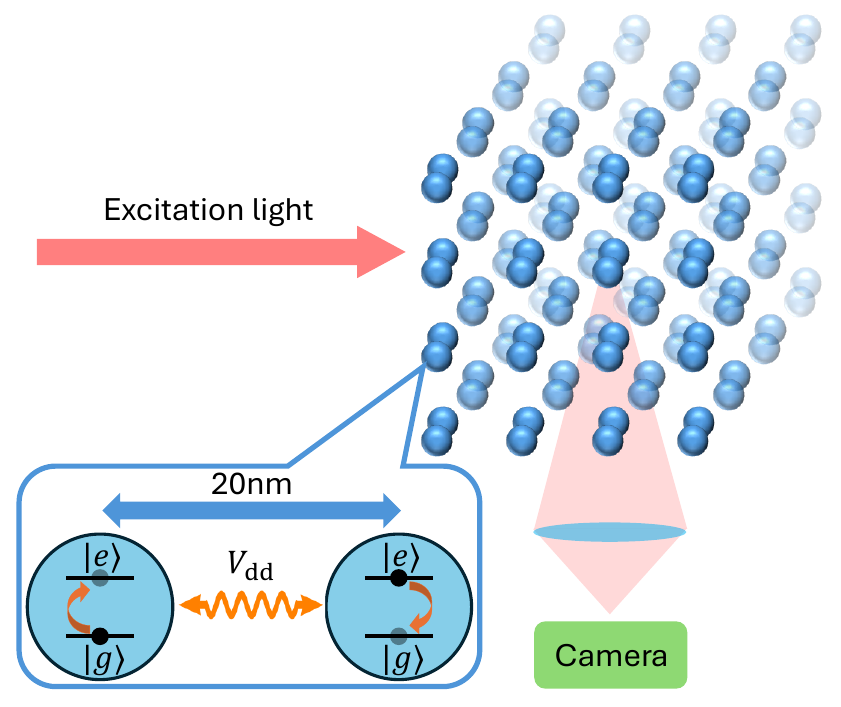}
    \caption[Atom pairs with strong dipole-dipole interactions.]{\textbf{Atom pairs with strong dipole-dipole interactions.} The atom pairs are confined to 20~nm size in a deep three-dimensional optical lattice. The laser beams for performing the resonant pulse or inducing the dipole moments are propagating along the horizontal and vertical directions, respectively (see main text), and the scattered light is collected from the bottom by a camera.}
    \label{fig:TwoAtom_Fig1}
\end{figure}

\begin{figure*}
    \centering
    \includegraphics[width=2\columnwidth]{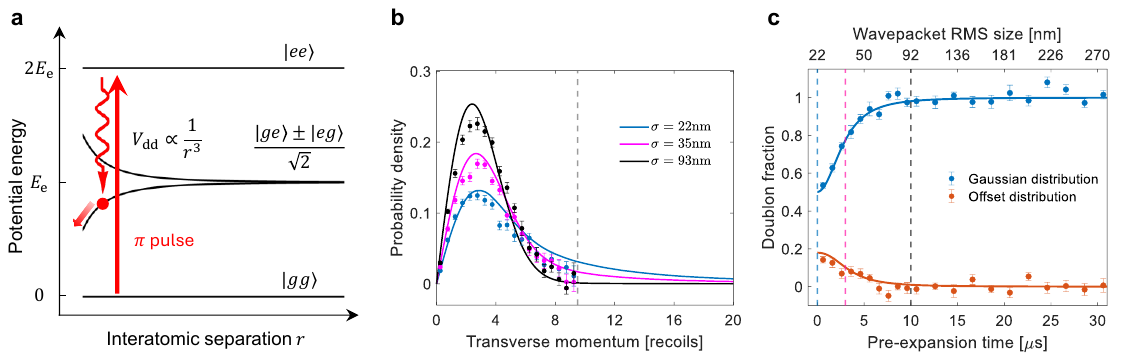}
    \caption[Large momentum transfer by one-photon emission.]{\textbf{Large momentum transfer by one photon emission.} \textbf{(a)}, Illustration of the large momentum transfer for atoms initialized in the single excitation manifold by a fast $\pi$-pulse and spontaneous emission of one photon. \textbf{(b)}, Observed momentum distributions after decay back to ground state for different initial wavepacket r.m.s. sizes 22nm (blue), 35nm (magenta), and 93nm (black). Solid lines represent simulations without any free parameter. The 3D median momenta are $7.8\hbar k$, $4.5\hbar k$, and $3.6\hbar k$ (see Methods). The dashed vertical line indicates the cutoff momentum for the detection region. \textbf{(c)}, The experimental data and numerical simulation of the atom number fraction for a bimodal fit (Gaussian plus an offset). The wavepacket size is tuned by expansion time before the laser pulse, and the blue, magenta, and black dashed lines correspond to expansion times in \textbf{(b)}. Each data point is averaged over independent experimental cycles, and the error bar reflects the s.e.m.
    }
    \label{fig:TwoAtom_Pi}
\end{figure*}

Surprisingly, few studies have been done on the light-induced electric dipole-dipole interaction between two atoms. Two atoms in well-defined quantum states are a paradigmatic and minimal framework for studying the electric dipole-dipole interaction between neutral atoms. Since the strength of the near-field dipole-dipole interaction is $\hbar\Gamma/ (r/\lambdabar)^3$, distances of less than 100 nm (corresponding to densities $n>10^{15}/\rm{cm^{3}}$) are required to make the coherent interaction strength larger than the spontaneous emission rate $\Gamma$. Here $\lambdabar=\lambda/2\pi$ is the reduced wavelength of the optical transition in the two-level atom. Such densities are challenging to achieve since they are accompanied by fast inelastic collisions.

In this work, we observed strong electric dipole-dipole interactions between two atoms by confining them to an r.m.s. size of about 20~nm on one site of an optical lattice, corresponding to a peak density of $ 2\times 10^{16}/\rm{cm}^3$. This extends the range of previous studies of cold atoms with $nk^{-3} \lesssim1$~\cite{PhysRevLett.113.133602,PhysRevLett.116.233601,PhysRevLett.116.183601,bromley2016collective,PhysRevLett.117.073003,PhysRevA.94.023842,PhysRevA.97.053816} by more than one order of magnitude to $nk^{-3} > 10$. Such densities are unprecedented in the field of ultracold atoms.
This new regime has allowed us to study several new phenomena of light-atom interactions by observing momentum transfer to atoms, scattered light, and atom loss: 
(1) For atom pairs in the electronically excited state, we demonstrate large momentum transfers $\sim10\hbar k$ associated with a single photon emission. This can be qualitatively understood by a single photon being exchanged between the two atoms multiple times before being emitted~\footnote{Since the photon exchange is virtual, this can lead to both repulsive and attractive forces}. (2) Far-detuned laser beams create atoms in dressed states with oscillating electric dipole moments of order 1 Debye.  We directly observe this via coherent light scattering and the strong mechanical effect of the atom pair as a novel coherent light-induced dipole-dipole interaction (which has been observed before only in its classical macroscopic limit). (3) In the fluorescence spectrum of atom pairs we observe a resonance shift due to the dipole-dipole interaction. We dub this effect as ``two-body Lamb shift'', analogous to the ``Lamb shift'' for single atoms interacting with vacuum modes and the ``collective Lamb shift'' for the interactions between many atoms. (4) Finally, we observed low light-induced (photoassociative) loss rates and a detuning asymmetry, which might be related to motional dynamics and complex excited state potentials. Most results are quantitatively explained by simple theoretical models. Note that we have limited our discussion to the optical domain. For long-wavelength transitions as between Rydberg states or polar molecules~\cite{carr2009cold}, the condition $nk^{-3}\gg1$ is easily fulfilled, but dissipative dynamics including spontaneous emission are much less important in those systems~\cite{browaeys2016experimental}.

\textbf{Atom pairs at high density}

To study the strong electric dipole-dipole interaction between two atoms, the atoms need to be positioned at close distances ($r\ll\lambdabar\approx100\rm{nm}$) which requires high atomic density ($n\gg10^{15}\rm{cm^{-3}}$). We fulfill this requirement by loading an almost pure BEC with around $5.5\times10^4$ of $^{162}\rm{Dy}$ atoms adiabatically into a deep three-dimensional (3D) optical lattice, with about $3\times10^4$ atoms in the $n=2$ Mott insulator core and $2.5\times10^4$ atoms in the $n=1$ Mott insulator shell (Fig.~\ref{fig:TwoAtom_Fig1}). The electronic ground state $\ket{J=8,m_J=-8}$ and the $\ket{J'=9, m_{J'}=-9}$ excited state of either the 421~nm or the 626~nm optical transition are connected via a cycling transition and form two-level systems, with the quantization axis along the vertical direction. Since there are only two atoms per site and they are prepared in the lowest spin state, there is no two-body or three-body loss, which makes it possible to reach densities $ > 10^{15}\rm{cm^{-3}}$. Once prepared in a superposition of ground and excited states, the atoms have dipole moments that rotate circularly in the horizontal plane. Our state preparation creates a cloud with doublons at the core and singlons in the outer Mott shell. We can obtain a doublon-only signal by recording a singlon-only signal (after removing the doublons), and subtracting it from the combined signal (see Methods). To control the strength of the dipole-dipole interaction, we change the size of the atomic wavepackets by either changing the optical lattice confinement or performing pre-expansion before measurements.

\textbf{Large momentum transfer by one-photon emission} 

\begin{figure*}
    \centering
    \includegraphics[width=2\columnwidth]{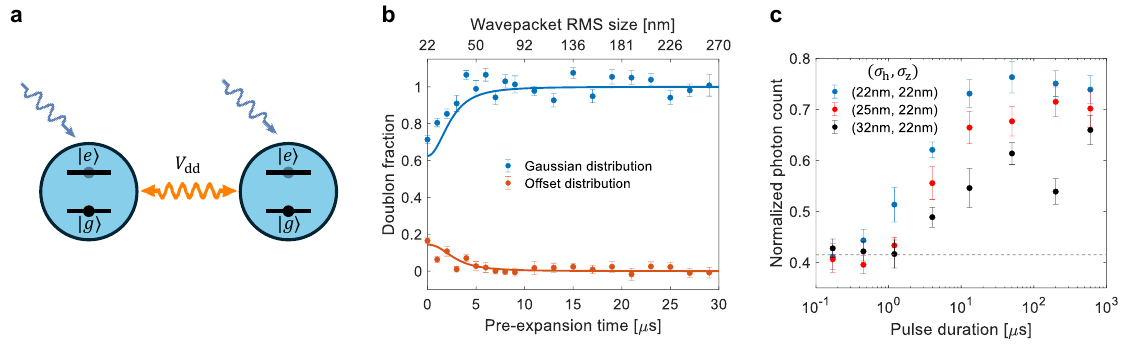}
    \caption[Coherent light-induced dipole-dipole interaction. ]{\textbf{Coherent light-induced dipole-dipole interaction.} \textbf{(a)}, Illustration of the light-induced electric dipole-dipole interaction between two atoms. \textbf{(b)}, The experimental data and numerical simulation of the atom number fractions in the two components of a bimodal fit to the time-of-flight momentum distribution. \textbf{(c)}, Normalized light scattering as a function of the pulse duration. The blue, red, and black data represent r.m.s. size in the horizontal direction of 22~nm, 25~nm and 32~nm. The detuning is 300~MHz blue detuned, and total light scattering for the fully expanded cloud is maintained at 0.3 photons scattered per atom by adjusting laser power accordingly. Each data point is averaged over independent experimental cycles, and the error bar reflects the s.e.m. }
    \label{fig:TwoAtom_LIDD}
\end{figure*}

Without interatomic interactions, the momentum transferred to the atom after one photon emission event is simply the recoil momentum $\hbar k$. However, the situation changes with interatomic interaction in the excited state manifold. For an atom pair initialized in the single excitation state, the electric dipole-dipole interaction causes strong acceleration before the atom pair decays to the ground state, and leads to a large momentum transfer (Fig.~\ref{fig:TwoAtom_Pi}a). For two-level atoms, the electric dipole-dipole interaction is $\sim\hbar\Gamma/(kr)^3$ in the near-field regime. The momentum transferred to the atom before the spontaneous emission can be estimated as $\Delta p=F\Delta t\sim3\hbar\Gamma/(k^3r^4)\times1/\Gamma=3\hbar k\times 1/(kr)^4$. Therefore, this momentum transfer can be much larger than the recoil momentum, given that the initial distance between the two atoms is much closer than $\lambdabar$~\footnote{This was qualitatively observed in a Ph.D. thesis~\cite{corman2016two}.}.

Experimentally, within 1 microsecond the lattice is switched off and the atom pairs are first prepared in the double-excitation state $\ket{ee}$ with a fast resonant $\pi$-pulse, then decay into the symmetric single-excitation state $(\ket{eg}+\ket{ge})/\sqrt{2}$ by the first spontaneous photon emission. Here, we choose the 626~nm optical transition because its relatively narrow linewidth ($\Gamma_{626}=2\pi\times 136\,\rm{kHz}$) allows pulses on an experimentally accessible time scale ($\sim50\,\rm{ns}$) before spontaneous emission happens. The electric dipole-dipole interaction in the symmetric single-excitation manifold leads to a large momentum transfer as shown in Fig.~\ref{fig:TwoAtom_Pi}b. For tight confinement (22 nm size)  the median momentum transfer between the atoms is about $8 \hbar k$ for each atom (Note that the average and rms momenta diverge for the case of a $1/r^3$ potential. In reality, the divergence is eliminated, since atom pairs at very close distance will not be excited because of the dipole-dipole shift and the finite Rabi frequency.). In this case, about 30\% of the atoms gain momentum larger than $10\hbar k$ and leave our detection region after ballistic expansion.  The momentum distribution evolves from a Gaussian to a bimodal distribution, which is well described by a Gaussian with an offset (see Methods). For larger wavepacket sizes, the averaged dipole-dipole interaction between the atom pair is reduced and so is the momentum transfer (Fig.~\ref{fig:TwoAtom_Pi}b). 
The measurements are done by switching off the lattice, optionally adding a first-stage expansion time to change the size of the  wavepackets  before applying the light pulse, and observing the momentum distribution  using an additional 4~ms time-of-flight. The experimental result of atom number under the narrow (Gaussian distribution) and broad (offset) distribution quantitatively matches the numerical simulations with no free parameters (Fig.~\ref{fig:TwoAtom_Pi}c). 

The physics of large momentum transfer after one photon emission has been studied in the context of atom loss caused by light-assisted collisions, where an atom gains enough kinetic energy from the electric dipole-dipole interaction to escape from the trap~\cite{RevModPhys.78.483}. Since the typical trap depth of optical tweezers is $\sim$mK (corresponding to momenta $\sim 100\hbar k$), and the trap depth for a magneto-optical trap is $\sim$K (corresponding to momenta $\sim 3000\hbar k$), the atom loss in previous studies of light-assisted collision occurs at the far tail of the momentum distribution and is a rare outcome during light scattering. In contrast, in our work, the momentum transfer per scattered photon is large. In addition, we can map out the dominant part of the momentum distribution of the atom pairs after one photon emission by time-of-flight imaging.

\textbf{Strong coherent light-induced dipole-dipole interaction}

So far, we have studied the dynamics of two atoms with a single excitation.  We now address another regime of strong dipolar forces, where each atom is continuously driven by the laser field and, as a dressed atom, acquires an (oscillating) electric dipole moment (approximately 1 Debye for our experimental parameters) which leads to strong dipolar interactions with other atoms (Fig.~\ref{fig:TwoAtom_LIDD}a).  These interactions are unitary and not directly related to photon emission: most of the dipolar force is experienced by the atoms between photon scattering events (see Methods). This light-induced dipole-dipole interaction has been discussed in numerous theoretical papers~\cite{PhysRevLett.84.5687,mazetsa2000ground,kurizki2002new,kurizki2004bose,low2005probing}.  The only experimental observation was in a BEC with 10,000 atoms where the effect of the oscillating dipoles is equivalent to a classical picture using the index of refraction of the cloud, which focuses the incident laser beam and generates a collective force~\cite{maiwoger2022observation}. Here we observe, for the first time, a strong laser-induced dipole-dipole force between an atom pair.

In the experiment, we used a $100\,\mathrm{ns}$ light pulse detuned by 16 linewidths to the blue side of the 421 nm resonance, during which about 0.25 photons are scattered per atom. We have also performed the same experiment on the opposite side of the resonance (16 linewidths red-detuned) or with different detunings (100 linewidths detuned), and obtained almost identical results, indicating the physics we study here is robust against the choice of detuning. The light-induced electric dipole force leads to a large momentum transfer between the two atoms due to a conservative force. In the Methods section we show that contributions to the mean force and momentum diffusion due to dissipation (photon scattering) become small for large detuning. Similar to the previous case with the $\pi$-pulse, we vary the wavepacket size by a first-stage expansion time. The momentum transfer broadens the original Gaussian momentum distribution to a bimodal shape, which is well captured by a sum of a Gaussian and an offset (see Methods). The data quantitatively agree with numerical simulations with no free parameters (Fig.~\ref{fig:TwoAtom_LIDD}b). The exact simulation can be well captured by a stationary model, and also by a simple analytic model that assumes atom pairs within a certain interatomic distance are ejected outside the narrow distribution (see Methods).

In addition to the mechanical effect, the electric dipole-dipole interaction also strongly modifies the light scattering of the atom pair. The dipole-dipole interaction shifts the atom pairs into resonance in the Franck-Condon (FC) region and enhances the photon scattering~\cite{RevModPhys.78.483}. The ratio of photon scattering from atoms in the FC region and the rest of the cloud is proportional to $n\lambdabar^3$ and is independent of laser detuning in the weak driving limit. Therefore, it is expected that the total photon scattering rate increases as the atom pair becomes more localized. Moreover, the extra light scattering in the FC region is largely incoherent due to the strong spatial variation of local detuning caused by the dipole-dipole shift, in contrast to the coherent off-resonant light scattering in the rest of the cloud (see Methods). 

Surprisingly, we have not observed this extra incoherent light scattering from the atoms in the FC region  (Fig.~\ref{fig:TwoAtom_LIDD}c). 
Instead, the normalized light scattering from the atomic array shows a similar suppression of light scattering as observed for singlons, for which the ratio of inelastic to coherent light is fully described by the Debye-Waller factor~\cite{lu2026suppression}. 

We attribute this finding to the motional dynamics of atoms during light scattering. For the detuning and intensity used in our experiment, atoms in the Franck-Condon region are resonantly excited and immediately ejected by the electric dipole-dipole interaction within a few ns. Even for the shortest pulse duration that we used, the population in the FC region quickly becomes almost completely depleted. This can be regarded as spatial (and spectral) hole burning in the pair wavefunction~\cite{PhysRevLett.126.083401}.

The observations of large momentum transfer (Fig.~\ref{fig:TwoAtom_LIDD}b) and  coherent light scattering (Fig.~\ref{fig:TwoAtom_LIDD}c) were done for the similar experimental conditions. This confirms that the off-resonant dipolar interactions are coherent and don't involve the open system dynamics of scattering photons into vacuum modes, in contrast to the previous $\pi$-pulse measurements.

We further study the coherence of the scattered light from the atom pairs by varying the light scattering pulse duration and the atomic wavepacket size (Fig.~\ref{fig:TwoAtom_LIDD}c).
The increase of light scattering for longer pulses is faster for tighter confinement and is possibly caused by partial refilling of the Franck-Condon region caused by the harmonic confinement. Note that for an unperturbed wavefunction (or using an infinite mass approximation), the normalized incoherent light scattering would be around 2, so our data suggest depletion of the Franck-Condon region by at least 50\% even for the longest and therefore weakest pulses.

\textbf{Observation of the two-body Lamb shift}

\begin{figure}
    \centering
    \includegraphics[width=\columnwidth]{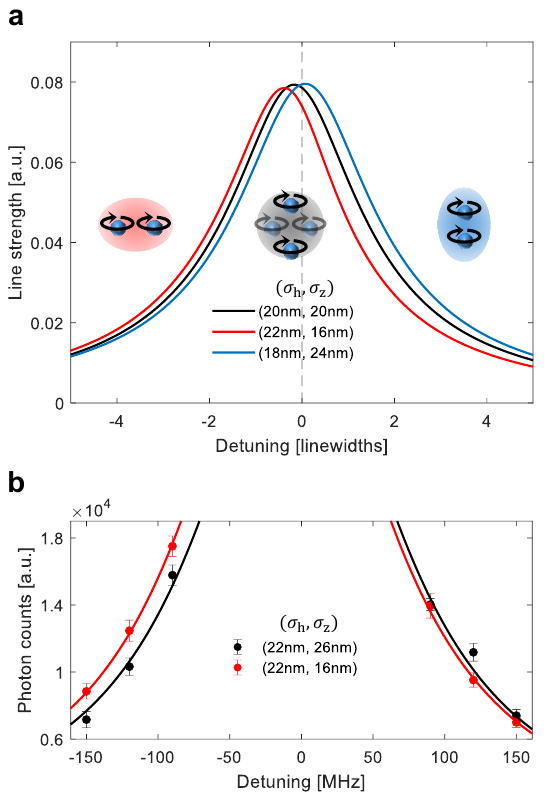}
    \caption[Observation of the two-body Lamb shift.]{\textbf{Observation of the two-body Lamb shift.} \textbf{(a)}, numerical simulation of the incoherent scattering spectrum under anisotropic confinement with different aspect ratios at constant density. A fit to the wings between $3\Gamma_{421}\leq|\Delta|\leq5\Gamma_{421}$ gives a shift of $0.54 \Gamma_{421}$ between the $(\sigma_\mathrm{h},\sigma_\mathrm{z})=(22~\mathrm{nm}, 16~\mathrm{nm})$ and $(22~\mathrm{nm}, 26~\mathrm{nm})$ cases in our experiment. \textbf{(b)}, spectroscopic measurement of the fluorescence spectrum of atom pairs under anisotropic confinement with different aspect ratios. Each data point is averaged over independent experimental cycles, and the error bar reflects the s.e.m. The solid lines are Lorentzian fits to the data and differ in center frequency by $8.6\pm2.5\mathrm{MHz}$ or $0.27(8)\Gamma_{421}$. This value is consistent with the theory prediction of 0.38 $\Gamma_{421}$ for the equal mixture of doublons and singlons in our sample (see Methods). 
     }
\label{fig:TwoAtom_Lamb}
\end{figure}

So far we have discussed that due to dipolar interactions, some parts of the wavefunction are on resonance, and other parts are off resonance with the laser light.  This should lead to a line broadening and a resonance shift when the laser frequency is scanned across the atomic resonance. The frequency shifts are large, on the order of $\Gamma/(k\sqrt{2}\sigma)^3\approx10 \Gamma$ for a wavepacket size of $\sigma\approx22\rm{nm}$. However, the anisotropic nature of the electric dipole-dipole interaction $V_{\rm{dd}}\propto(3\cos^2\theta-1)/r^3$ ($\theta$ being the angle between the quantization axis and the interatomic axis) reduces the interaction strength dramatically once averaged over all orientations. We call the remaining energy shift the two-body Lamb shift.

Indeed, for a spherically symmetric wavepacket, the resonance shift of the fluorescence spectrum is almost zero. The average dipolar interaction energy in a spherical cloud is exactly zero, but the convolution with the Lorentzian spectrum gives a small spectral shift.  To achieve an observable shift,  the aspect ratio of the wavepacket needs to be modified: when the wavepacket size is smaller in the vertical direction, the two atoms are more likely to align side-by-side, which results in an attractive dipole-dipole interaction and a red shift of the resonance. The opposite shift occurs for prolate wavepackets (Fig.~\ref{fig:TwoAtom_Lamb} a). 

Experimentally, the fluorescence spectrum of the atom pairs are measured by using the 421~nm optical transition of dysprosium, for different r.m.s. size of the wavepacket in the vertical direction and a fixed horizontal r.m.s. size of 22~nm. We chose the 421~nm transition because its broad linewidth ($\Gamma_{\rm{421}}=2\pi\times 32\,\rm{MHz}$) makes the measurement insensitive to light shifts produced by the lattice beams, although the two-body Lamb shift in units of the linewidth ($\sim n\lambdabar^3$) is smaller than for the 626~nm transition. We used weak intensity (resonant saturation parameter around $s\approx2\times10^{-4}$) and long duration (0.5ms) to minimize the mechanical effect of the dipole-dipole interaction.

Observing shifts in dense clouds is difficult due to strong absorption and possible lensing effects and collective interactions~\cite{PhysRevA.96.053629,1wb7-1z1b}. Our typical optical density is around 20. Since we are interested in the two-body Lamb shift, which depends only on the pair wavefunction, we can discriminate against other effects by modifying the pair wavefunction without altering the geometry of the entire cloud.

In Fig.~\ref{fig:TwoAtom_Lamb} b we have varied the aspect ratio of the two-body wavefunction and compared the fluorescence spectrum for ($\sigma_{\rm{h}}=22\,\rm{nm}$, $\sigma_{\rm{z}}=26\,\rm{nm}$) and ($\sigma_{\rm{h}}=22\,\rm{nm}$, $\sigma_{\rm{z}}=16\,\rm{nm}$). The first case is consistent with no shift, while the second case shows a red shift of $0.27(8)\Gamma_{421}$. This observation establishes the two-body Lamb shift. 

Because of limitations in laser power for the lattice beams, we could not access the opposite aspect ratio with sufficiently deep lattices, necessary for a significant blue shift. The observed resonance shift of $8.6\pm2.5\mathrm{MHz}$ agrees with the modeling in Fig.~\ref{fig:TwoAtom_Lamb}a semi-quantitatively. Our simple model (see Methods) assumes that the wavefunction does not change during the light pulse, which becomes exact in the perturbative limit.

We performed various tests to rule out other explanations.  Light shifts due to the lattice beams were ruled out by varying the lattice depth for singlons without observing any shift. Clouds after short time-of-flight of 300~$\mathrm{\mu s}$ have similar optical density, but did not show any shift of the resonance. In the perturbative regime, the photoassociative loss during the light pulse is negligible, and does not depend on the sign of the detuning, as confirmed by the fluorescence atom count after the light pulse. Therefore, we also rule out the possibility of spectrum asymmetry induced by detuning-dependent losses.

There is the possibility of collective Lamb shifts due to the $1/r$ retarded dipole-dipole interactions.  These shifts can depend on size and shape of the localized wavepackets. In the Methods section we show theoretically and experimentally that the many-body Lamb shift is affected by the horizontal confinement, but not by the vertical confinement. This provides strong evidence that the shift observed in Fig.~\ref{fig:TwoAtom_Lamb}b (by varying the vertical confinement) is due to the dipolar interactions within the pairs of atoms.

Note that for a linewidth much smaller than the motional state splitting the spectrum is simply a  Lorentzian shifted by the dipolar energy averaged over the ground state wave function~\cite{PhysRevLett.123.223601,PhysRevA.101.043816}.  Here, we are in the opposite limit where each detuning creates an excitation mainly in the narrow Franck-Condon region which implies dipolar mixing of the motional states.

\textbf{New regime for photoassociation}

\begin{figure}
    \centering
    \includegraphics[width=\columnwidth]{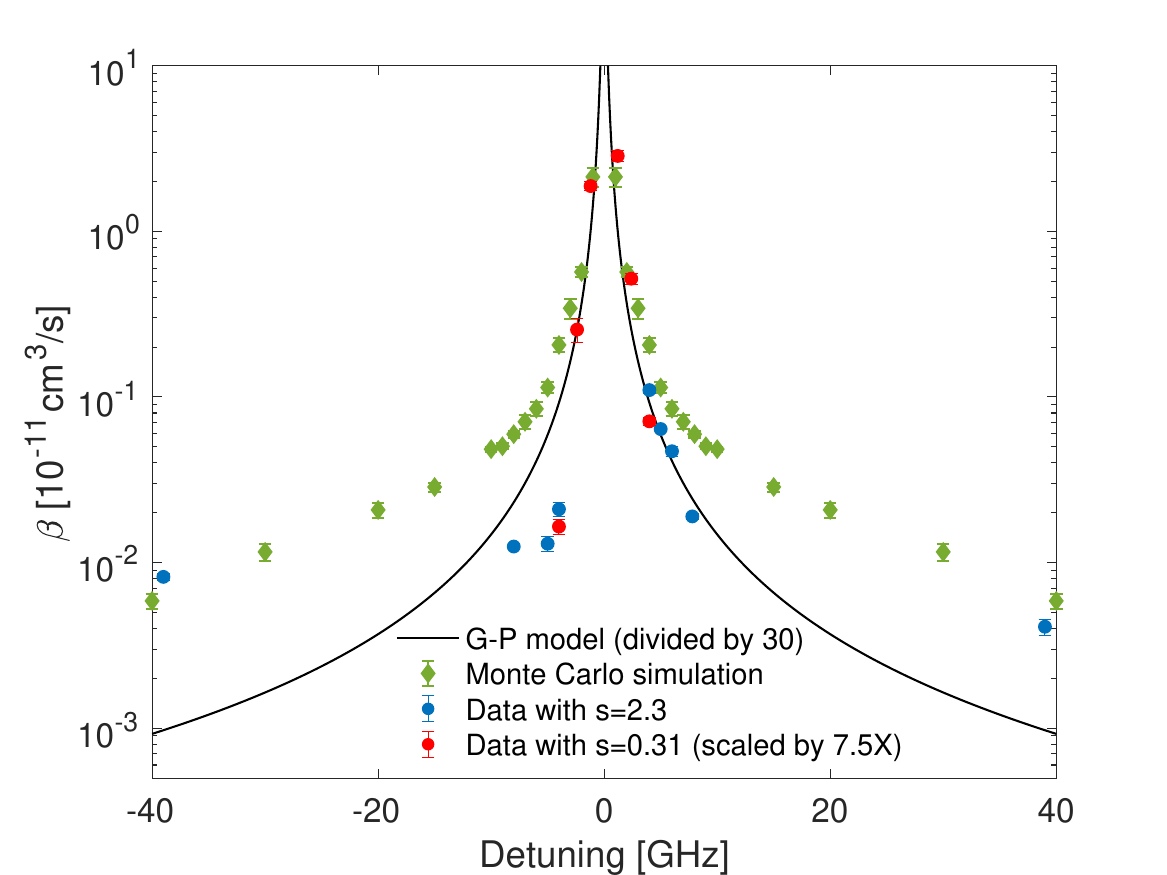}
    \caption[Photoassociation spectroscopy for atom pairs.]{\textbf{Photoassociation spectroscopy for atom pairs.} The two-body loss rate coefficient $\beta$ of photoassociation loss is measured as a function of detuning for the 421nm optical transition. The Gallagher-Pritchard (G-P) model prediction is scaled down by a factor of 30 to guide the eye, and the data with lower power $s\approx0.31$ is scaled by 7.5 to normalize for the power difference. Each data point is averaged over independent experimental cycles, and the error bar reflects the s.e.m. Over the range of the measurement, the Landau-Zener parameter $\pi\hbar\Omega^2/|v_{\mathrm{rel}}U'(R_c)|$ ranges from 2.1 at 4~GHz to 0.10 at 39~GHz.  }
    \label{fig:TwoAtom_PA}
\end{figure}

Our measurements of light scattering have shown that the Franck-Condon region gets depleted by the strong dipolar forces in the excited states.
Since it is the Franck-Condon region where photoassociative loss occurs, we expect to see strong correlations between photo-induced losses and fluorescence measurements.

We want to point out that atoms prepared in an $n=2$ Mott insulator are an ideal platform to study photoassociation~\cite{PhysRevLett.93.073002,PhysRevLett.108.173002,PhysRevLett.134.013202, RevModPhys.78.483}. The atoms are in a well-defined quantum state, and there is no competing three-body loss. Furthermore, at the high densities possible ($>10^{16}/\rm{cm}^3$) even very weak photoassociation processes can be measured, which can be relevant to characterize molecular potentials.

Experimentally, we illuminate the atom pairs in an optical lattice with 421~nm laser light near the saturation intensity ($s\approx2.3$ and $s\approx0.31$) of the transition and vary the detunings. The remaining atom number is measured as a function of time, from which the effective two-body loss rate coefficient $\beta$ is extracted (see Methods). The results are summarized in Fig.~\ref{fig:TwoAtom_PA}. 
In general, we observe loss rate coefficients 5 to 40 times smaller than predictions of the Gallagher-Pritchard (G-P) model. 
In the range 1-5 GHz, we observe an additional suppression of loss by up to a factor of 5 on the red detuning side. A similar detuning asymmetry was observed in the coherence of light scattering.
Both photoassociative loss and incoherent resonant light scattering occur in the Franck-Condon region. Both kinds of observations reveal depletion of the Franck-Condon region which (for reasons which we don't fully understand) is not symmetric in laser detuning.  Due to motional dynamics, the Franck-Condon region of the relative wavefunction is depleted even for very small values of the Landau-Zener parameter. The observed lower loss rate compared to the Gallagher-Pritchard model is semi-quantitatively explained by semi-classical simulations (see Methods).

Regarding the detuning asymmetry of $\beta$, models assuming a non-interacting ground state wavefunction predict similar result for positive and negative detuning. Although the solid angle for attractive interactions in the dipolar potential is larger than for repulsive interactions, this is exactly compensated by the relative strength of the interactions, and the effective volume of the Franck-Condon regions is the same. This applies to both light scattering and the Gallagher-Pritchard model, and the classical Monte-Carlo simulations for photoassociation.  
Previous measurements for Rb showed an asymmetry between red and blue detuning of at most a factor of two~\cite{PhysRevLett.134.013202}.

Strong asymmetry can  come from bound molecular states, which usually appear at red detunings of 300 linewidths~\cite{PhysRevLett.93.123202}. Although no such resonance was observed in our measurements, we speculate that the observed asymmetry is related to the unresolved structures of the complex molecular potentials of dysprosium~\footnote{Note that trap loss measurements in very deep traps can also show an asymmetry, since for blue detuned light, the maximum energy transfer to the two atoms is limited to the detuning. We are in the limit of shallow traps, where excitation in the FC region always leads to trap loss.}.

\textbf{Discussion and outlook}

In conclusion, we have studied a new regime for strong electric dipole-dipole interactions between tightly confined atom pairs. The dipole-dipole interaction leads to large momentum transfer in single spontaneous photon emission and in coherent light-induced dipole-dipole interactions. It also leads to a two-body Lamb shift when the atom pair is prepared in an anisotropic wavefunction. In this new regime, both light scattering and photoassociation are profoundly modified by the dynamics of the pair wavefunction.

Compared to previous studies on atom pairs in light fields, we have prepared the atoms in a single quantum state and can trigger strong dipole-dipole interactions with $\mu$s resolution, and observe momentum distributions, coherence of light scattering, and losses.  Our findings strongly suggest further studies on the quantum dynamics of photoassociation, especially since this process is an important technique for preparing single atoms for quantum computation~\cite{fung2015efficient,grunzweig2010near,brown2019gray,jenkins2022ytterbium,aliyu2021d,ang2022gray,zhu2025high}.

Our studies of strong dipolar interactions can be extended in many ways. The two-body Lamb shift would be much larger if two atoms were localized at sub-50 nm distances using our recently demonstrated sub-wavelength control of Dy atoms~\cite{du2024atomic}. In this case, the Lamb shift would not be reduced by averaging over all angles of orientation, and it should be possible to resolve the splitting between the symmetric and anti-symmetric states.

Our studies of dipolar forces and dynamic correlations can be extended to atoms prepared in a Mott insulator with three atoms per site, ultimately connecting the few-body and many-body regimes. One should also explore light scattering by atom pairs for other species. The wavefunction dynamics and the fluorescence spectrum will depend on the mass of the atoms and the linewidth of the optical transitions.   Finally, the light-induced dipole-dipole interaction has been suggested as a way to engineer new types of interactions (e.g. $1/r$ gravity-like potentials~\cite{PhysRevLett.84.5687,mazetsa2000ground,kurizki2002new,kurizki2004bose,low2005probing}) and our work is a first step into these directions.

\emph{Note}.---Related work with a much lighter atom ($^7$Li) has explored other aspects of two-atom light scattering including collective enhancement of coherent scattering and modifications due to the scattering length~\cite{Li7Doublon}.

\noindent\textbf{Acknowledgments}.---We are grateful to Ana Maria Rey, Hanzhen Lin, and David Pritchard for discussions. 
We thank William R. Milner for proofreading the manuscript. We acknowledge support from the NSF (grant No. PHY-2208004), from the Center for Ultracold Atoms (an NSF Physics Frontiers Center, grant No. PHY-2317134), from the Vannevar-Bush Faculty Fellowship (grant No. N00014-23-1-2873), from the Gordon and Betty Moore Foundation (GBMF ID \# 12405), from the Army Research Office (contract No. W911NF2410218) and from the Defense Advanced Research Projects Agency (award HR0011-23-2-0038). Yu-Kun Lu is supported by the NTT Research Fellowship. Jiahao Lyu is supported by the MathWorks Science Fellowship.

\nocite{corman2016two}
\bibliography{Ref.bib}

\newpage
\clearpage

\setcounter{figure}{0}
\setcounter{equation}{0}
\makeatletter 
\renewcommand{\thefigure}{S\@arabic\c@figure}
\renewcommand{\theequation}{S\@arabic\c@equation}
\renewcommand{\theHfigure}{S\arabic{figure}}
\renewcommand{\theHequation}{S\arabic{equation}}
\makeatother


\section*{Methods}
\subsection{Experimental procedures}

\subsubsection{Sample preparation}

We prepare a 3D Mott insulator by adiabatically loading $^{162}$Dy ground state atoms (in the $|J=8,m_J=-8\rangle$ state) from an almost pure BEC into a 3D optical lattice. The rectangular lattice comprises two horizontal 1,064 nm retro-reflected beams with  $80$ degrees intersection angle, and one vertical 532 nm retro-reflected beam. At the end of the evaporative cooling process, we turn on a magnetic field gradient to compensate 90\% of gravity, which allows us to reduce the trap frequency of the optical dipole trap during the lattice loading process and ensures a larger doublon portion of the Mott insulator.

The lattice ramp consists of two steps. In the first step, we exponentially ramp up the three lattices to $(30 \,E_\mathrm{R,h},30\,E_\mathrm{R,h},30\,E_\mathrm{R,v})$ in 370 ms, where recoil energies $E_\mathrm{R,h}=h\times1.09\, \mathrm{kHz}$, $E_\mathrm{R,v}=h\times4.35\, \mathrm{kHz}$ for $^{162}$Dy. At this moment, the atoms are deeply in the Mott insulator regime, where the only relevant energy scale is the lattice band gap, on the order of 10 kHz. Then we quickly ramp the lattices linearly to their final depths in 1~ms.

After loading the atoms into the Mott insulator state, the lattice filling is determined in the following way. In a deep lattice $(900 \,E_\mathrm{R,h},900\,E_\mathrm{R,h},200\,E_\mathrm{R,v})$, the triplons will be lost in 1~ms because of the high atomic density~\cite{PhysRevA.109.023319}. We have verified that $4.5\times 10^4$ atoms can be held in an $(900 \,E_\mathrm{R,h},900\,E_\mathrm{R,h},200\,E_\mathrm{R,v})$ optical lattice without any loss for 20~ms, showing that there is no triplon shell  for this atom number. Note that our studies of doublons are robust against a small fraction of triplons. The doublon fraction is determined by inducing pairwise loss caused by dipolar relaxation after two rapid adiabatic passage (RAP) radiofrequency (RF) sweeps spaced by 5~ms hold time. The singlons return to the initial state after the two RAP sweeps, while the doublons are lost during the hold time after the first RAP sweep. For all experiments, the number of atoms on singlon sites and doublon sites are approximately equal.

\subsubsection{Probe beam and imaging description}
We use probe laser beams near the $^{162}$Dy 421~nm and 626~nm transitions for different measurements in this paper. The 421~nm probe beam in the Lamb shift measurement and the 626~nm probe beam are in the same plane as the horizontal lattice beams, propagating at an angle of approximately 36$^{\circ}$ with respect to one of the horizontal lattice beams. 
Both probe beams have horizontal linear polarization, while the $^{162}$Dy spin always points in the vertical direction. Due to the large difference in the Clebsch-Gordan coefficients for $\sigma^+$ and $\sigma^-$ transitions, the $^{162}$Dy atoms in state $|J=8,m_J=-8\rangle$ predominantly interact with the $\sigma^-$ component of the probe beam. Therefore, we can safely consider $^{162}$Dy atom as a two-level system which consists of $|g\rangle  \equiv |J=8,m_J=-8\rangle$ and $|e\rangle \equiv |J'=9,m_{J'}=-9\rangle$ interacting with the probe beams in all cases considered in this paper. 

For the $\pi$-pulse experiment and the data shown in Fig.~\ref{fig:TwoAtom_Pi}c, we use a resonant highly saturated laser beam at 626 nm propagating horizontally with linear polarization in the horizontal plane to invert the two-level system population in about 50 ns, much faster than  the excited state lifetime $\tau_{626}=1.2$ $\mu$s.

For the light-induced dipole-dipole and the photoassociation measurement, we used $\sigma^-$ polarized light propagating along the vertical direction which drives the $ |J=8,m_J=-8\rangle$ to $|J'=9,m_{J'}=-9\rangle$ cycling transition.

The imaging in this paper was done by either fluorescence or absorption imaging. For the light scattering measurement in Fig.~\ref{fig:TwoAtom_LIDD}c and the Lamb shift measurement, we use fluorescence imaging along the vertical direction that collects photons emitted downwards with a 3-inch diameter NA=0.5 aspheric lens, similar to our previous work~\cite{lu2026suppression}. For all other measurements, we use absorption imaging along the horizontal probe beam direction.

\begin{figure}
    \centering
    \includegraphics[width=\linewidth]{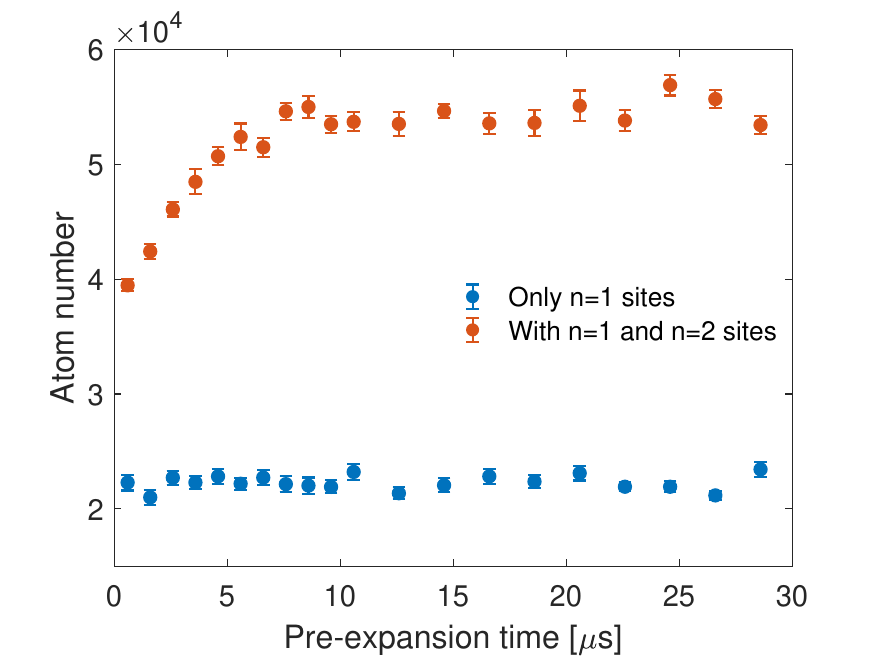}
    \caption{\textbf{Raw data for the dipolar force measurements.} For the measurement of dipolar force after the optical $\pi$-pulse (Fig.~\ref{fig:TwoAtom_Pi}c), we compare the atom number in the narrow momentum distribution with (red) and without doublons (blue). Each data point is averaged over independent experimental cycles, and the error bar reflects the s.e.m. No apparent atom loss is observed for $n=1$ filling, whereas up to 50\% of atoms are lost when doublon sites are filled. Identical procedure is applied for the light-induced dipole force measurement in Fig.~\ref{fig:TwoAtom_LIDD}b.}
    \label{fig:n1}
\end{figure}

\subsubsection{Determination of doublon contribution}
For all the data presented in Fig.~\ref{fig:TwoAtom_Pi}c and Fig.~\ref{fig:TwoAtom_LIDD}b, the contribution of the $n=1$ shell is determined  by removing the doublons via dipolar relaxation before the light scattering pulse and subtracting the signal from the total signal (i.e. without doublon removal). We conducted an identical set of experiments where only singlons were prepared and found no dependence of atom count on wavepacket size (Fig.~\ref{fig:n1}).

\subsubsection{Determination of the momentum distribution of atom pairs}

\begin{figure}
    \centering
    \includegraphics[width=\linewidth]{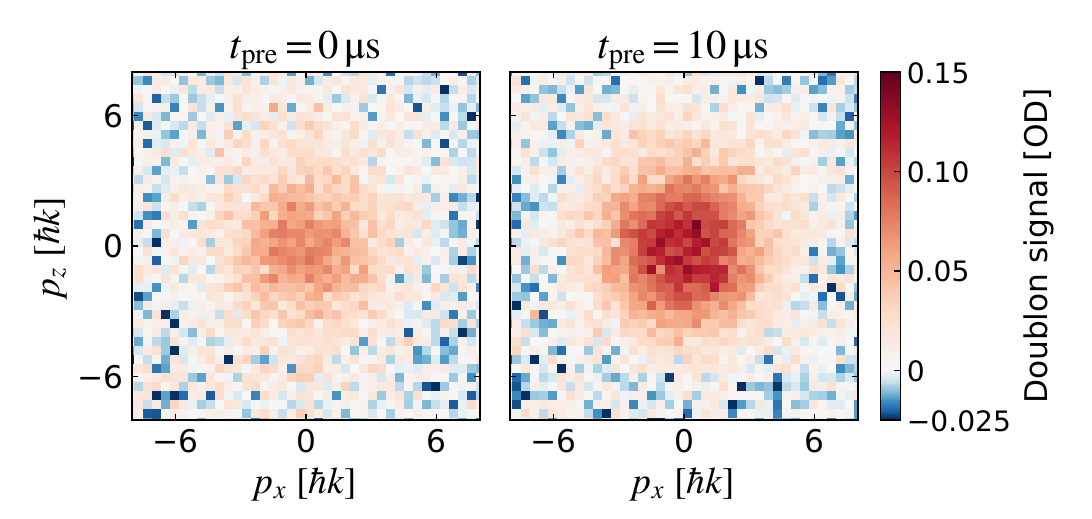}
    \caption{\textbf{Momentum distribution of atom pairs after an optical $\pi$-pulse.} Averaged absorption images of doublons acquired after the 626 nm $\pi$-pulse and 4~ms of time of flight are shown for pre-expansion times \(t_{\mathrm{pre}}=0\,\mu\mathrm{s}\) (left) and \(10\,\mu\mathrm{s}\) (right). The images used \(5\times5\) pixel binning for display. Momenta are expressed in units of the 626~nm photon recoil \(\hbar k\). The central doublon signal is weaker at \(0\,\mu\mathrm{s}\) compared to \(10\,\mu\mathrm{s}\) because the dipolar interaction pushes atoms to higher momentum for smaller initial interatomic distance.}
    \label{fig:TOFimage}
\end{figure}

The transverse momentum distribution of the atom pairs in Fig.~\ref{fig:TwoAtom_Pi}b was obtained by the following procedure. At each pre-expansion time, we averaged the absorption images of the singlon-doublon mixture and singlon-only preparations separately and subtracted the singlon-averaged image (Fig.~\ref{fig:TOFimage}). For each preparation condition, we repeated the experiment 12 times and averaged the images. We then summed the resulting doublon signal in 0.5-recoil annuli up to 9.5 recoils and divided by the bin width and the initial doublon-atom population to obtain the momentum distribution in Fig.~\ref{fig:TwoAtom_Pi}b. The analysis was repeated on images resampled with replacement separately for each pre-expansion time. Error bars show the standard deviation across bootstrap replicates in each momentum bin. 

\subsection{Modeling the mechanical effects of dipole-dipole interaction}

\begin{figure*}
    \centering
    \includegraphics[width=1.0\linewidth]{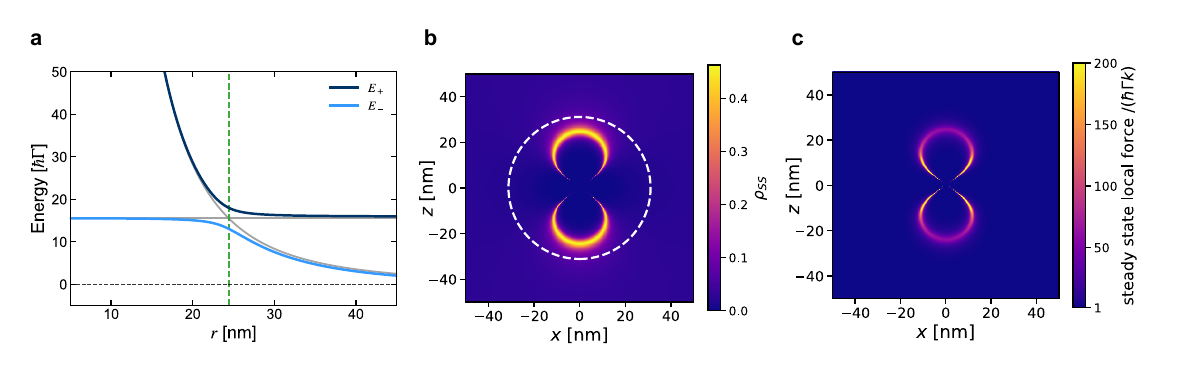}
    \caption{\textbf{Excited state fraction and dipolar forces during light scattering.} \textbf{a}, Avoided crossing energy diagram for $\Delta= 2\pi\times500$ MHz, $\Omega/\Gamma=3.5$. \textbf{b}, The excited state fraction $\rho_{ SS}(\boldsymbol{r})$ peaks in the Franck-Condon region. The Franck-Condon region has varying thickness and radius, depending on the relative angle $\theta$ between the two atoms. The white circle shows the relative r.m.s. size of the atom pair. \textbf{c}, Steady-state local force $F(\boldsymbol{r})$ (Eq.~\ref{eq:S7}) also peaks in the Franck-Condon region.} 
    \label{fig:Condon}
\end{figure*}

The simulations used in Fig.~\ref{fig:TwoAtom_Pi} and Fig.~\ref{fig:TwoAtom_LIDD} are inspired by the Gallagher-Pritchard (G-P) model and the treatment of light-assisted collision with optical Bloch equations~\cite{RevModPhys.71.1}. In short, we treat the internal state dynamics of the atom pair quantum mechanically and  the external motion classically. To simulate the mechanical effect caused by either the 626~nm $\pi$-pulse or 421~nm light pulse, we developed a Monte Carlo simulation to find the classical trajectory of the atom pair. 

We consider two atoms A and B, initialized in their ground state $|g_\mathrm{A}g_\mathrm{B}\rangle$. Their relative distance $|\boldsymbol{r}|\equiv|r_\mathrm{A}-r_\mathrm{B}|\ll \lambdabar$, with $\lambdabar$ being the reduced wavelength of a given optical transition, but larger than the $s$-wave scattering length $\sim$ 7~nm. 

In the $\pi$-pulse experiment (Fig.~\ref{fig:TwoAtom_Pi}), the pair of atoms is quickly excited to $|e_\mathrm{A}e_\mathrm{B}\rangle$. After a time $\sim1/(2\Gamma_{626})$ it decays to the fully symmetric state $|S\rangle=\frac{1}{\sqrt{2}}(|e_\mathrm{A}g_\mathrm{B}\rangle +|g_\mathrm{A}e_\mathrm{B}\rangle)$ and begins to interact via the strong dipolar forces. The lifetime of the symmetric state is reduced by a factor of 2 due to the superradiance effect, after which the pair decays to its ground state and flies away freely. In the case of the continuously driven 421 nm transition (Fig.~\ref{fig:TwoAtom_LIDD}), the atom pair internal wavefunction quickly reaches the optical Bloch equation steady-state solution $\rho=\rho_\mathrm{ss,AB}(\boldsymbol{r})$ after $t\approx 1/(2\Gamma_{421}) \approx 2.5 \, \rm{ns}$, and the dipole interaction stays on until the 100~ns pulse ends. Because of the tight localization of the atoms, we neglect the relative propagation phase difference between the two oscillating dipoles, so the atom pair is always in the fully symmetric state. The detailed expression for the density matrix $\rho$ will be explained later. We want to first point out that the steady state density matrix strongly depends on the relative position of the two dipoles, because of the $1/r^3$ potential experienced by the atom pair in the laser field~\cite{Gallagher1989}. 

The electric dipole-dipole Hamiltonian between a pair of two-level atoms in the near field reads~\cite{PhysRevA.47.1336}:

\begin{equation}
\begin{aligned}
     H_{\mathrm{dd}}(\boldsymbol{r})&=\frac{1}{4\pi \epsilon_0 r^3}\left(\boldsymbol{\hat{d}}_\mathrm{A}\cdot \boldsymbol{\hat{d}}_\mathrm{B}-3(\boldsymbol{\hat{d}}_\mathrm{A} \cdot \boldsymbol{u})(\boldsymbol{\hat{d}}_\mathrm{B}\cdot \boldsymbol{u})\right)\\
     &=\frac{3\hbar \Gamma(3\cos^2{\theta}-1)}{8(kr)^3}(\hat{\sigma}_{+}^\mathrm{(A)}\hat{\sigma}_{-}^\mathrm{(B)}+\mathrm{h.c.}),
\end{aligned}
    \label{equation:S1}
\end{equation}
where we define the relative position $\boldsymbol{r}=\boldsymbol{r}_\mathrm{A}-\boldsymbol{r}_\mathrm{B}=r\boldsymbol{u}$, $\boldsymbol{u}$ being a unit vector pointing from $\mathrm{B}$ to $\mathrm{A}$. The Pauli matrix is $\hat{\sigma}_+ \equiv |e\rangle\langle g|$, and $\theta$ is the angle between $\boldsymbol{u}$ and the $z$-axis.

In classical electrodynamics \cite{jackson1998classical}, Eq.~\ref{equation:S1} is equivalent to the interaction energy of two oscillating electrical dipoles with dipole moments $\boldsymbol{d}=\langle \psi | \hat{\boldsymbol{d}}|\psi\rangle$, if $|\Psi\rangle_{AB}=|\psi\rangle_A|\psi\rangle_B$. For the $\pi-$pulse case, the internal state of $|S\rangle$ is fully entangled, and one has to use the quantum picture to get the correct interaction energy, which is twice larger than the classical one.

Since size of the atomic wavefunction is much smaller than $\lambdabar$, we have restricted ourselves to the symmetric subspace: $\{|G\rangle \equiv |g_\mathrm{A}g_\mathrm{B}\rangle ,|S\rangle \equiv \frac{1}{\sqrt{2}}(|g_\mathrm{A}e_\mathrm{B}\rangle+|e_\mathrm{A}g_\mathrm{B}\rangle),\, |E\rangle \equiv |e_\mathrm{A}e_\mathrm{B}\rangle \}$. In this basis, the dipole-dipole interaction (DDI) is diagonal, and only the state $|S\rangle$ has a non-zero energy shift 
\begin{equation}
    E_{S}(\boldsymbol{r})=\frac{3\hbar \Gamma}{8(kr)^3}(3\cos^2{\theta}-1).
    \label{eq:S2}
\end{equation}

We use Eq.~\ref{eq:S2} to simulate both the 626~nm $\pi-$pulse and the continuous 421~nm pulse experiment. In both simulations, our initial atom pair position and momentum $(\boldsymbol{r}_\mathrm{A},\boldsymbol{p}_\mathrm{A},\boldsymbol{r}_\mathrm{B},\boldsymbol{p}_\mathrm{B})$ are randomly chosen from a normal distribution corresponding to the Wigner distribution of a quantum harmonic oscillator ground state after a variable pre-expansion time. 

The simulation proceeds through discrete time with step size $\Delta t =10^{-4}/\Gamma_\mathrm{626}$ for the $\pi-$pulse experiment, and $\Delta t =5\times10^{-2}/\Gamma_\mathrm{421}$ for the continuous 421~nm probe experiment. For the $\pi-$pulse experiment, the atom pair is first excited to state $\ket{E}$ before decaying to the symmetric state $\ket{S}$. We ignore the Van der Waals interaction in state $\ket{E}$ since it is much weaker than the dipole-dipole interaction. Therefore, we use $|S\rangle$ as the pair initial internal state, and perform a classical simulation of the trajectories of $\sim10^5$ atom pairs under the potential given by Eq.~\ref{eq:S2}. After each time step, the pair has a small probability $2\Gamma_{626} \Delta t$ of returning to the ground state $|G\rangle$. After the photon emission event, the pair will keep constant velocities. The single-photon kick $\hbar k$ for excitation and spontaneous emission is small compared to the momentum transfer by the dipole-dipole interaction and is therefore neglected. To avoid divergence of the interacting potential, we include a hard-core potential with a $7 \,\mathrm{nm}$ radius. 

For the continuous 421 nm probe experiment, the atomic external and internal wavefunctions entangle in a more subtle way. To proceed with the semi-classical simulation, we need to derive a position-dependent averaged force $F = F(\boldsymbol{r})$ in order to simulate the classical motion of the atom pair.

In the symmetric subspace, the system Hamiltonian is given by 
\begin{equation}
    H=T+ H_{\mathrm{drive}}+H_{\mathrm{int}}
\end{equation}

The three terms are kinetic energy, single atoms coupled to the probe laser field, and DDI term respectively.

\begin{equation}
\begin{cases}
    T=(\frac{\hat{\boldsymbol{p}}_A^{2}}{2m}+\frac{\hat{\boldsymbol{p}}_B^{2}}{2m})\otimes \mathbb{I}_{\mathrm{s}}\\
    H_{\mathrm{drive}}=\mathbb{I}_{\mathrm{mot}}\otimes\hbar[\frac{\Omega}{2} (\hat{S}^-+\hat{S}^+) - \Delta |S\rangle\langle S| - 2\Delta |E\rangle\langle E|]\\
    H_{\mathrm{int}}(\boldsymbol{r})=E_S(\boldsymbol{r})\mathbb{I}_{\mathrm{mot}}\otimes |S\rangle\langle S| ,
\end{cases}
\end{equation}
where $\Delta$ is the laser detuning away from the bare resonance, $\Omega$ is the single atom Rabi oscillation frequency, and $\hat{S}^-(\hat{S}^+)$ is collective spin lowering(raising) operator. We ignore the interaction between two ground or two excited-state atoms, since they are much weaker than the light-induced dipole-dipole interactions. The kinetic energy term $T$ in the Hamiltonian can be treated as a perturbation if the mass $m$ is large. In the infinite mass limit, we can solve the Schr\"{o}dinger equation or Lindblad master equation for each relative position. The equation can be solved either in a two-level system $\{ |G\rangle,\,|S\rangle \}$ if we ignore the population in $|E\rangle$ in the weak drive limit, or in a full three-level system. For our experimental parameters, the two simulations give almost identical results. Thereafter, we discuss the physics in a two-level description, although the full numerics was done for the three-level model.

With a non-zero laser drive $\Omega$, $\{|G\rangle,|S\rangle\}$ have an avoided crossing at the Condon point where $\Delta=E_S(\boldsymbol{R}_{c})/\hbar$ with a gap of $\sqrt{2}\Omega$. At a far distance, the asymptotic eigenstates are $|G\rangle,|S\rangle$ with an energy splitting of the laser detuning $\Delta$. Fig.~\ref{fig:Condon}a
 shows the 1D cut of the eigenenergies as a function of interatomic distance.

Including the dissipative part, the full Lindblad master equation reads:
\begin{equation}
    \frac{\mathrm{d} \rho}{\mathrm{d}t}=-\frac{i}{\hbar}[H,\rho]+\Gamma\left(\hat{S}^-\rho \hat{S}^+-\frac12\{\hat{S}^+\hat{S}^-,\rho\}\right)
\end{equation}

The steady-state solution in the infinite-mass (or stationary) approximation is given by the optical Bloch equations, with the laser detuning modified by the local DDI:

\begin{equation}
\begin{cases}
    \rho_{SS}(\boldsymbol{r}) = \frac{1}{2} \Omega^2 \left[ \delta(\boldsymbol{r})^2 + \Omega^2 + \Gamma^2 \right]^{-1} \\[1ex]
    \Re(\rho_{GS}(\boldsymbol{r})) =\frac{\sqrt2}{2} \Omega \delta(\boldsymbol{r}) \left[ \delta(\boldsymbol{r})^2 + \Gamma^2 + \Omega^2 \right]^{-1} \\[1ex]
    \Im(\rho_{GS}(\boldsymbol{r})) = \frac{\sqrt2}{2}\Omega \Gamma \left[ \delta(\boldsymbol{r})^2 + \Gamma^2 + \Omega^2 \right]^{-1}\\[1ex]
    \rho_{GG}(\boldsymbol{r})=1-\rho_{SS}(\boldsymbol{r}).
\end{cases}
\label{eq:S6}
\end{equation}
Here $\delta(\boldsymbol{r}) = \Delta - E_S(\boldsymbol{r})/\hbar$ is the local detuning. The usefulness of Eq.~\ref{eq:S6} will be shown later. Note that we mainly care about the terms diagonal in position space, where we define $\rho_{\alpha \beta}(\boldsymbol{r})\equiv \langle \boldsymbol{r},\alpha|\rho|\boldsymbol{r},\beta\rangle$. This Born-Oppenheimer type adiabatic approximation is valid when the atomic motion during the relaxation time $1/(2\Gamma_{421})$ of the optical Bloch equation (OBE) is negligible. The atom moves $\Delta x\sim F/(4m\Gamma_{421}^2)$ during the OBE time constant $1/(2\Gamma_{421})$, with $F$ being the dipole-dipole force experienced by the atom. The corresponding change of the local detuning $\delta(\boldsymbol{r})$ over distance $\Delta x$ in units of the linewidth is $\Delta\delta(\boldsymbol{r})/(2\Gamma_{421})=\Delta x |\nabla\delta(\boldsymbol{r})|/(2\Gamma_{421})$, and takes the maximum value $\sim9 E_\mathrm{R}/(4\hbar\Gamma_{421}(kr_c)^8)$ near the Condon point at $r=r_c$. Since $\Delta=\Gamma_{421}/(kr_c)^3$, we can re-express $\Delta\delta(\boldsymbol{r})/(2\Gamma_{421})$ using the detuning as $\Delta\delta(\boldsymbol{r})/(2\Gamma_{421})=9E_\mathrm{R}(\Delta/\Gamma_{421})^{8/3}/(4\hbar\Gamma_{421})$.

For $\Delta\sim4\Gamma_{421}$ in the Lamb shift measurement, $\Delta\delta(\boldsymbol{r})/(2\Gamma_{421})\sim0.02$. For $\Delta\sim10\Gamma_{421}$ in the light-induced dipole-dipole measurement, $\Delta\delta(\boldsymbol{r})/(2\Gamma_{421})\sim0.2$. For $\Delta\sim100\Gamma_{421}$ in the photoassociation measurement, $\Delta\delta(\boldsymbol{r})/(2\Gamma_{421})\sim100$. Note that the estimation above is for atoms near the Condon point, while most of the atoms are far away, where the dipole-dipole force and the atomic motion are much smaller. Therefore, in the light-induced dipole-dipole experiment, the internal spin degree of freedom of the atom pair is mostly in the steady state of the optical Bloch equations during the whole classical trajectory. In contrast, for the photoassociation experiments, the atom pair crosses the Franck-Condon region before decaying back to the ground state and can gain substantial kinetic energy. For light atoms such as Li~\cite{Li7Doublon}, $\Delta\delta(\boldsymbol{r})/\Gamma$ is much higher.

Note that the assumption that the atom pair is fully excited to state $\ket{S}$ with a probability that depends on the atom pair separation (as used in Ref.~\cite{grun2025light}) would incorrectly create large momentum kicks for each scattered photon and would fail to explain the observed coherent light scattering (see main text). 

Once the steady-state population for each position is obtained, we can calculate the averaged force in the Heisenberg picture:

\begin{equation}
    F(\boldsymbol{r})=\frac{\mathrm{d} \langle \boldsymbol{p} \rangle }{\mathrm{d} t}=-\langle \frac{\partial H}{\partial \boldsymbol{r}} \rangle = -\langle\rho_{SS} \frac{\partial E_S}{\partial \boldsymbol{r}}\rangle.
    \label{eq:S7}
\end{equation}

\begin{figure}
    \centering
    \includegraphics[width=\linewidth]{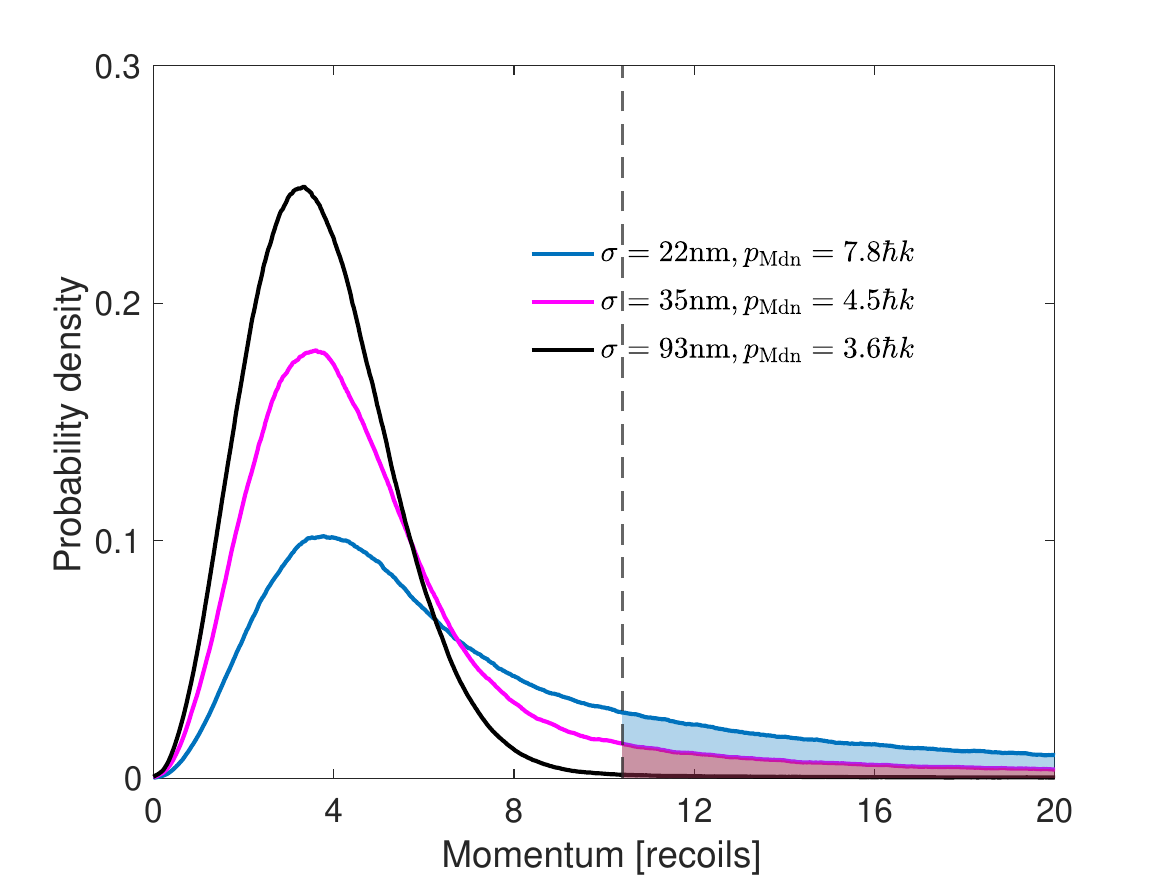}
    \caption{\textbf{Simulated 3D momentum distribution of the atom pairs after the $\pi$-pulse.} 3D Momentum distribution in Fig.~\ref{fig:TwoAtom_Pi} after decay back to the ground state for different initial wavepacket r.m.s. sizes 22nm (blue), 35nm (magenta), and 93nm (black). The 3D median momenta are $7.8\hbar k$, $4.5\hbar k$, and $3.6\hbar k$. The dashed vertical line indicates the cutoff momentum for the detection region.}
    \label{fig:3Ddistribution}
\end{figure}

\begin{figure}
    \centering
    \includegraphics[width=\linewidth]{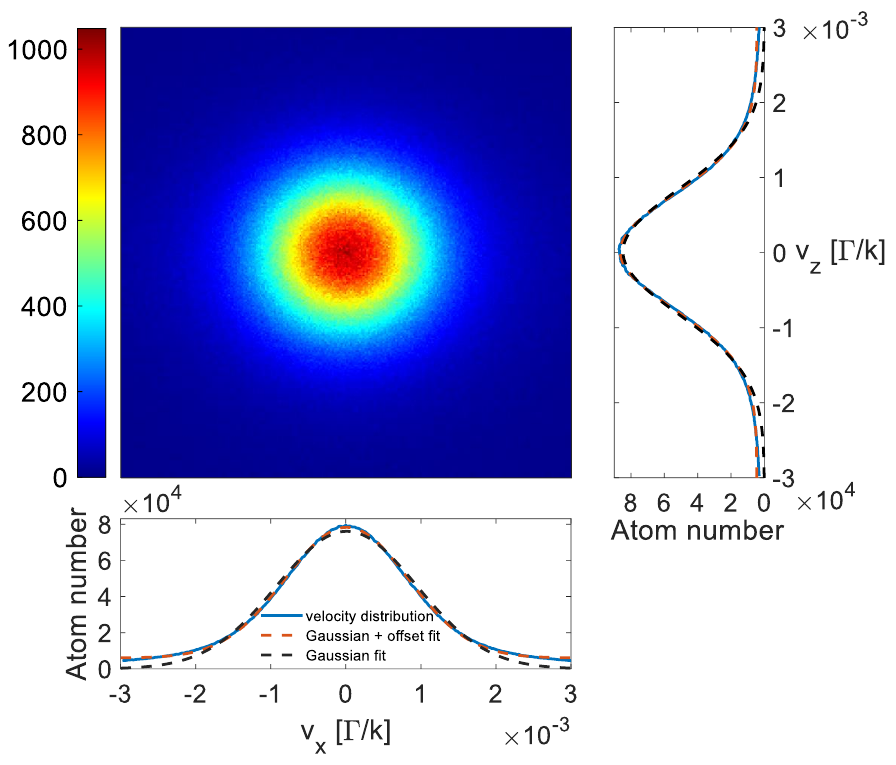}
    \caption{\textbf{Momentum distribution simulation result.} The momentum distribution after a 100~ns light pulse with 500~MHz detuning from the 421 nm line is simulated and integrated along our imaging direction $y$ (the 2D plot). The momentum distribution can be further integrated along either $x$ or $z$ direction (the 1D plots), and can be well fitted by a Gaussian distribution with an offset, but not by a plain Gaussian distribution. This fitting result establishes the bimodal nature of the momentum distribution. }
    \label{fig:2Dmomentum}
\end{figure}

For our experimental parameters, an example of the steady-state excited state population and total force is shown in Fig.~\ref{fig:Condon}b and Fig.~\ref{fig:Condon}c. The simulated momentum distribution is non-Gaussian because the dipolar interaction pushes atoms at close interatomic distance to high momenta (Fig.~\ref{fig:3Ddistribution}), but can be represented by a Gaussian with an offset (Fig.~\ref{fig:2Dmomentum}). This ansatz was also used to fit the experimental data of the time-of-flight images.

\vspace{20pt}

\textbf{Derivation of the classical momentum width result from a quantum picture.} 
Here we use a toy model to show  how a classical model of the momentum distribution emerges from the full quantum picture. In the toy model, we suppose the atom pair's relative wavefunction is frozen during a short interaction time $\tau$ and accumulate phase under the influence of the interaction $V(\boldsymbol{r})$, which is assumed to have even parity. In the Schr\"{o}dinger picture, this process appears as $\psi(r)\rightarrow\psi(r)e^{-iV(\boldsymbol{r})\tau/\hbar}$. In the Heisenberg picture, the wavefunction remains Gaussian, but the momentum operator changes: $\hat{p}\rightarrow (\hat{p}+\hat{F}(\boldsymbol{r})\tau)$. In the classical case, the momentum for a particle initally at $(\boldsymbol{r,p})$ becomes $(\boldsymbol{r},\boldsymbol{p}+F(\boldsymbol{r})\tau)$.  Also note that the momentum after the pulse is an odd function, and the atom distribution in both the quantum and the classical case are even. Therefore, all the odd moments of the distribution are zero. The time-of-flight picture captures the lowest order moments of the momentum, especially the variance of the momentum spread, which we will prove to be equal in both cases. 

For the classical case, the average is taken over the whole phase space with uncorrelated position and momentum: $\rho_{cl}(\boldsymbol{r},\boldsymbol{p}) = \rho_r(\boldsymbol{r})\rho_p(\boldsymbol{p})\propto \exp(-r^2/2\sigma_r^2-p^2/2\sigma_p^2)$. The variance in the classical case is $\langle (\boldsymbol{p}+F(\boldsymbol{r})\tau)^2\rangle_{cl}=\langle p^2 \rangle_{cl}+\langle F(\boldsymbol{r})^2\rangle_{cl}\tau^2 $. The cross term cancels, since the position and momentum are uncorrelated, and $\langle p\rangle=0$. 

In the quantum case, the operator average is taken over the Gaussian wavefunction. The average of the cross term $\langle \hat{p}\hat{F}(\boldsymbol{r})+\hat{F}(\boldsymbol{r})\hat{p}\rangle_{q}$ is also zero for a real-value Gaussian wavepacket, because the two terms are purely imaginary and conjugate to each other. In the end, $\langle(\hat{p}+\hat{F}(\boldsymbol{r})\tau)^2\rangle_q=\langle\hat{p}^2\rangle_q+\langle \hat{F}(\boldsymbol{r})^2\rangle_q\tau^2$. Now, if we choose the classical Wigner distribution according to the momentum and position uncertainty of the coherent wavepacket, the two results will be exactly the same.

\textbf{Stationary model for the momentum distribution.}

Because of the heavy mass and broad linewidth of the dysprosium atom, the atomic motion is negligible during the 100~ns 421~nm light pulse.  This allows a simple description of the momentum distribution caused by the light-induced dipole-dipole interaction in a stationary model where the light-induced dipole-dipole force is constant and the momentum transfer depends only on the initial displacement between the atom pair: $\Delta p=F(\boldsymbol{r})\Delta t$. The momentum distribution given by this stationary model matches the exact simulation result perfectly.

Finally, there is a simple analytic way to describe the fraction of atoms that receive a large momentum transfer and are missing from the narrow distribution in Fig.~\ref{fig:TwoAtom_LIDD}. We assume that only atom pairs with an initial distance below a threshold $r_0$ are ejected out of the narrow distribution. By integrating the Gaussian relative density distribution above $r=r_0$, we obtain $\text{erfc}(\frac{r_0}{\sqrt{2}\sigma})+\sqrt{\frac{2}{\pi}}\frac{r_0}{\sigma}e^{-\frac{r_0^2}{2\sigma^2}}$. 
This result reproduces the narrow distribution in Fig.~\ref{fig:TwoAtom_LIDD}b, if we choose  $r_0=0.61\lambdabar$ as a fitting parameter.

\subsection{Coherent and incoherent light scattering of the atom pair}

\begin{figure}
    \centering
    \includegraphics[width=\linewidth]{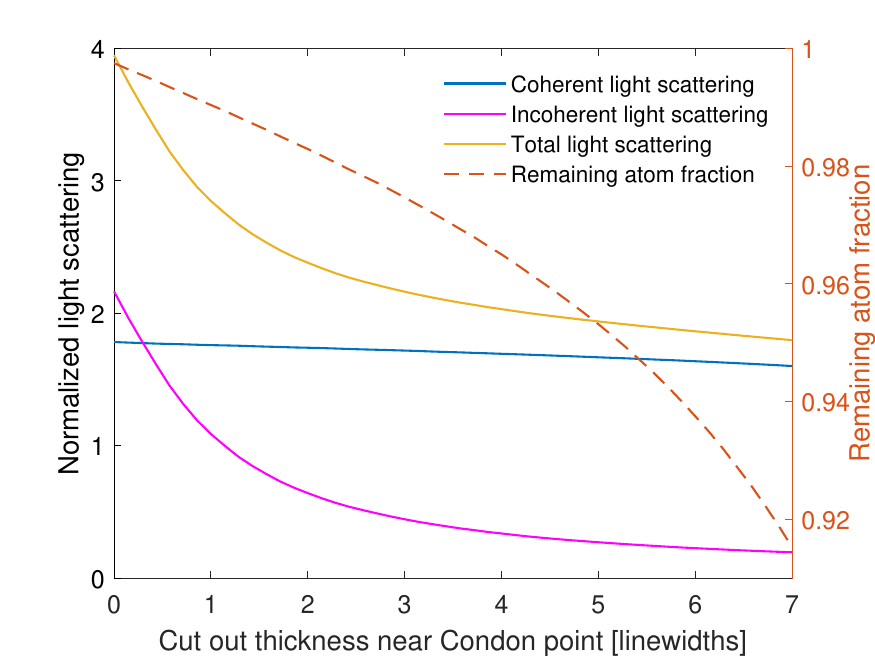}
    \caption{\textbf{Light scattering by eliminating the Franck-Condon region from the cloud.}  This model reflects that the atoms in the Franck-Condon region are depleted by the dipolar force. Cutting out the Franck-Condon region leads to a large degree of coherent light scattering.  At angle $\theta=0^\circ$ near the Condon point, the local detuning shift by about 1 linewidth in 1~nm. We assumed 20~nm r.m.s. size of atomic density in all directions and 10 $\Gamma$ detuning for the calculation here.}
    \label{fig:Cutoff}
\end{figure}

In this section, we explain the formalism to calculate the coherent and incoherent light scattering of the atom pair. In a simplified model, we take the mass to be infinite.  Therefore, atomic motion is absent and the ground-state Gaussian relative density distribution $n(\boldsymbol{r})$ of the atom pair is unmodified. When the laser drives the atoms continuously, the steady-state density matrix can be written as
\begin{equation}
    \langle \boldsymbol{r}|\rho|\boldsymbol{r}\rangle=
    \begin{bmatrix}
        \rho_{GG}(\boldsymbol{r)}&\rho_{GS}(\boldsymbol{r})\\
        \rho_{SG}(\boldsymbol{r})&\rho_{SS}(\boldsymbol{r})
    \end{bmatrix}n(\boldsymbol{r}),
\end{equation}
where the internal state density matrix is given by Eq.~\ref{eq:S6}. The off-diagonal matrix elements in real space are decoupled from the diagonal elements and do not contribute to the observables we need later. Also, since we use a weak probe to measure the Lamb shift (resonant saturation parameter $s\approx 2\times 10^{-4}$), we can safely eliminate the doubly excited state $|E\rangle=|e_\mathrm{A}e_\mathrm{B}\rangle$ and work in the $\{|G\rangle,|S\rangle\}$ subspace. 

In the Heisenberg picture, the electric field operator of the scattered light is directly related to the dipole moment of the atoms by $E^+(t)\propto \hat{S}_-(t)=\int d\boldsymbol{r}\hat{S}_-(t)\otimes\left|\boldsymbol{r}\rangle\langle \boldsymbol{r} \right|$. The coherent and total scattering intensities are given by $I_\text{coh}(t)=\left|\braket{E^+(t)}\right|^2\propto \left|\int d\boldsymbol{r} \rho_{GS}(\boldsymbol{r})n(\boldsymbol{r})\right|^2$ and $I_\text{total}(t)=\braket{E^-(t)E^+(t)}\propto\int d\boldsymbol{r} \rho_{SS}(\boldsymbol{r})n(\boldsymbol{r})$, respectively~\cite{lu2026suppression}.

The dipole moment reaches a steady state and oscillates at the driving frequency, but the position-dependent detuning causes a position-dependent amplitude and phase of the induced dipole moment, which causes incoherent light scattering. Note that in our model, all the phase factors associated with atom position are taken to be zero, which implies a Debye-Waller factor of 1~\cite{zwhd-1k2t}. This assumption is made to be self-consistent with the symmetric subspace we have chosen. The Debye-Waller factor for our highly localized wavepackets (Fig.~\ref{fig:TwoAtom_Lamb}) is about 0.85. In fact, we will show that the dominant incoherence comes from position-dependent detuning $\delta(\boldsymbol{r})$.

To obtain a physical picture of this incoherent light scattering, we show below that the physics can be understood by an effective momentum transfer of the photon scattering caused by the dipole-dipole interaction. If we turn on the laser to drive the atoms continuously, the system will reach the steady state with excited state fraction $\braket{\ket{S}\bra{S}}=\rho_{SS}(x,t)n(x)=\frac{\Omega^2/2}{\delta(x)^2+\Omega^2+\Gamma^2}n(x)$ and dipole moment $d(x,t)=\braket{\ket{G}\bra{S}}=\frac{\Omega(\delta(x)+i\Gamma)/\sqrt{2}}{\delta(x)^2+\Omega^2+\Gamma^2}n(x)$. The total scattering is $I_\text{total}(t)=\braket{E^-(t)E^+(t)}\propto\int \rho_{SS}(x,t)n(x)dx=\int dx \frac{\Omega^2/2}{\delta(x)^2+\Omega^2+\Gamma^2}n(x)$, while the coherent scattering is $I_\text{coh}(t)=\left|\braket{E^+(t)}\right|^2\propto \left|\int d(x,t)dx\right|^2=\left|\int\frac{\Omega(\delta(x)+i\Gamma)/\sqrt{2}}{\delta(x)^2+\Omega^2+\Gamma^2}n(x)dx\right|^2$. In this case, the dipole moment reaches the steady state and oscillates at the driving frequency so there is no extra dephasing from the local detuning, but the position-dependent detuning causes a position-dependent complex amplitude which also causes incoherent scattering.

\subsection{Momentum transfer due to dipolar interactions
}
\label{subsec:scattering_momentum}

In this section, we derive the momentum transfer caused by the dipolar interaction between two atoms. We assume a one-dimensional model in which the two-level system consisting of $\ket{G}$ and $\ket{S}$ is being driven by a laser field with Rabi frequency $\sqrt{2}\Omega$ and a spatially varying detuning $\delta(x)=\Delta-E_S(x)/\hbar$ caused by the dipolar interactions. The excited state linewidth is $2\Gamma$ due to superradiance.

In the weak excitation limit $\Omega^2\ll\delta^2+\Gamma^2$ and assuming slow atomic motion, we can adiabatically eliminate the excited state and obtain an additional light-shift potential and an effective quantum jump operator for the ground state~\cite{PhysRevA.85.032111}:
\begin{equation}
\begin{aligned}
    U_{\mathrm{coh}}(x)
    &=\frac{\hbar\Omega^2\delta}{2(\delta^2+\Gamma^2)},\\
    L(x)&=\frac{\sqrt{\Gamma}\,\Omega}
    {\delta(x)+i\Gamma}.
\end{aligned}
\label{eq:scat_effective_operators}
\end{equation}
Note that we neglect the single-photon recoil $k$ here. 
The motional density operator $\rho$ therefore evolves as
\begin{equation}
\begin{aligned}
    \dot\rho&=-\frac{i}{\hbar}[H_0+U_{\mathrm{coh}},\rho]
    +\mathcal D[L]\rho,\\
    \mathcal D[L]\rho&=L\rho L^\dagger-\frac12\{L^\dagger L,\rho\}.
\end{aligned}
    \label{eq:scat_master}
\end{equation}

The effective Hamiltonian produces the coherent dipole force
$F_{\mathrm{coh}}=-U_{\mathrm{coh}}'$, and the quantum jump operator is responsible for the dissipative force
$F_{\mathrm{dis}}=\hbar\,\mathrm{Im}(L^\dagger L')$.
Writing $L=\sqrt{R}\,e^{i\phi}$, where
$R=|L|^2=\Gamma\Omega^2/(\delta^2+\Gamma^2)$ is the local scattering rate, we obtain
\begin{equation}
\begin{aligned}
    F_{\mathrm{coh}}
    &=\frac{\Omega^2}{2(\delta^2+\Gamma^2)^2}
    \left(\delta^2-\Gamma^2\right)F_S,\\
    F_{\mathrm{dis}}
    &=\hbar R\phi'
    =\frac{\Omega^2\Gamma^2}{(\delta^2+\Gamma^2)^2}F_S,\\
    F_{\mathrm{coh}}+F_{\mathrm{dis}}
    &=\frac{\Omega^2}{2(\delta^2+\Gamma^2)}F_S
    =\rho_{SS}F_S.
\end{aligned}
\label{eq:scat_mean_force}
\end{equation}
Here $\rho_{SS}=\Omega^2/\left(2(\delta^2+\Gamma^2)\right)$ is the excited-state population
to leading order in the drive, and $F_S=\hbar \delta'$ is the full dipole force in the excited state $\ket{S}$. The two contributions therefore recover
the averaged dipole force we derived before. The dissipative drift is set by the phase gradient $\phi'$, and is much smaller than the coherent dipole force when $|\delta|\gg\Gamma$.

The quantum jump operator also produces momentum diffusion. The time evolution of $p^2$ is given by

\begin{equation}
\begin{aligned}
    \frac{dp^2}{dt}
    &=\{p,F_{\mathrm{coh}}+F_{{\mathrm{dis}}}\}+B_{pp},\\
    B_{pp}
    &=\hbar^2|L'|^2=R\,\frac{F_S^2}{\delta^2+\Gamma^2}.
\end{aligned}
\label{eq:scat_diffusion}
\end{equation}

Here $B_{pp}$ characterizes the momentum diffusion on top of the mean momentum drift, and has contributions from both the amplitude and phase gradients of $L$. The mean dissipative momentum transfer per photon scattering event and the
corresponding r.m.s. momentum noise are given by
\begin{equation}
\begin{aligned}
    \frac{F_{\mathrm{dis}}}{R}
    &=\frac{\Gamma F_S}{\delta^2+\Gamma^2},\\
    p_{\mathrm{rms}}
    &\equiv\sqrt{\frac{B_{pp}}{R}}
    =\frac{|F_S|}{\sqrt {\delta^2+\Gamma^2}}.
\end{aligned}
    \label{eq:scat_momentum_scales}
\end{equation}
The latter characterizes ensemble momentum diffusion and does not
represent a fixed momentum kick at each emission. 

In the limit $\delta \gg \Gamma$, the coherent force is larger than the dissipative force by a factor of $\delta^2/(2\Gamma^2)$, and the normalized coherent force $F_{\mathrm{coh}}/R$ is larger than $p_{\mathrm{rms}}$ by a factor of $\delta/(2\Gamma)$. This shows quantitatively how the dipolar interaction becomes a conservative force for large detuning.

Far from resonance,
$p_{\mathrm{rms}}\simeq|F_S|/|\delta|$  is dominated by the amplitude
gradient of $L$. Note that the momentum diffusion of an atom with spatially varying detuning can be calculated exactly beyond the weak driving limit~\cite{GATheory} by generalizing the treatment for atoms in standing waves~\cite{PhysRevA.21.1606}.

The quantum jump operators determine the coherence of the emitted
light. In the weak-excitation limit, the coherent and total intensities
in the channel considered here satisfy
$I_{\mathrm{coh}}\propto|\langle L\rangle|^2$ and
$I_{\mathrm{tot}}\propto\langle L^\dagger L\rangle$, with averages over
the motional state. For a Gaussian density distribution centered at
$x_0$, with r.m.s.\ width $\sigma$, and a scattering amplitude that varies
weakly across the distribution one obtains
\begin{equation}
\begin{aligned}
    \frac{I_{\mathrm{coh}}}{I_{\mathrm{tot}}}
    &=1-\sigma^2\left|\frac{L'(x_0)}{L(x_0)}\right|^2
    +O(\sigma^4)\\
    &=1-\sigma^2 k_{\mathrm{eff}}^2+O(\sigma^4),\\
    k_{\mathrm{eff}}
    &\equiv\left|\frac{L'(x_0)}{L(x_0)}\right|
    =\frac{|\delta'(x_0)|}{\sqrt{\delta^2(x_0)+\Gamma^2}}\\
    &=\frac{p_{\mathrm{rms}}(x_0)}{\hbar}.
\end{aligned}
\label{eq:scat_coherence}
\end{equation}
The effective wavevector thus connects momentum
diffusion to the reduction of coherent light scattering. For a pure recoil operator $L\propto e^{iqx}$, the coherent light scattering fraction reduces to the usual Debye--Waller factor $e^{-q^2\sigma^2}$; a position-dependent scattering amplitude generally
requires the full ratio $|\langle L\rangle|^2/\langle L^\dagger L\rangle$.

As a result, off-resonant scattering has an effective r.m.s. momentum transfer $\hbar k_{\rm{eff}}\sim  |F_S(x_0)/\delta|$ per photon or the photon recoil, whichever is larger. The coherent scattering fraction of the atom pair is then given by the effective Debye–Waller factor $D_{\rm{eff}}=\left|\braket{\exp{\left(-ik_{\rm{eff}}x\right)}}\right|^2=\exp{(-k_{\rm{eff}}^2\sigma^2)}$ which is close to unity.  This implies a small degree of entanglement between photon and atom, and suppression of light scattering in periodic arrays~\cite{lu2026suppression}. 

Applying this model to our experiments, the photon scattering inside (outside) the FC region is associated with a large (small) momentum transfer and is mainly incoherent (coherent).
For the initial Gaussian relative density distribution between the two atoms, most of the incoherent light scattering comes from atoms in the FC region. But during the light pulse, the atoms in the FC region feels a strong force and are quickly ejected.  This lowers the incoherent scattering substantially but keeps the coherent scattering almost unchanged (Fig. \ref{fig:Cutoff}).

\subsection{Modeling of the Lamb shift}
This section discusses the simulation related to the fluorescence spectrum of the atom pair in Fig.~\ref{fig:TwoAtom_Lamb}. 
\subsubsection{Modeling of the two-body Lamb shift}

\begin{figure}
    \centering
    \includegraphics[width=\linewidth]{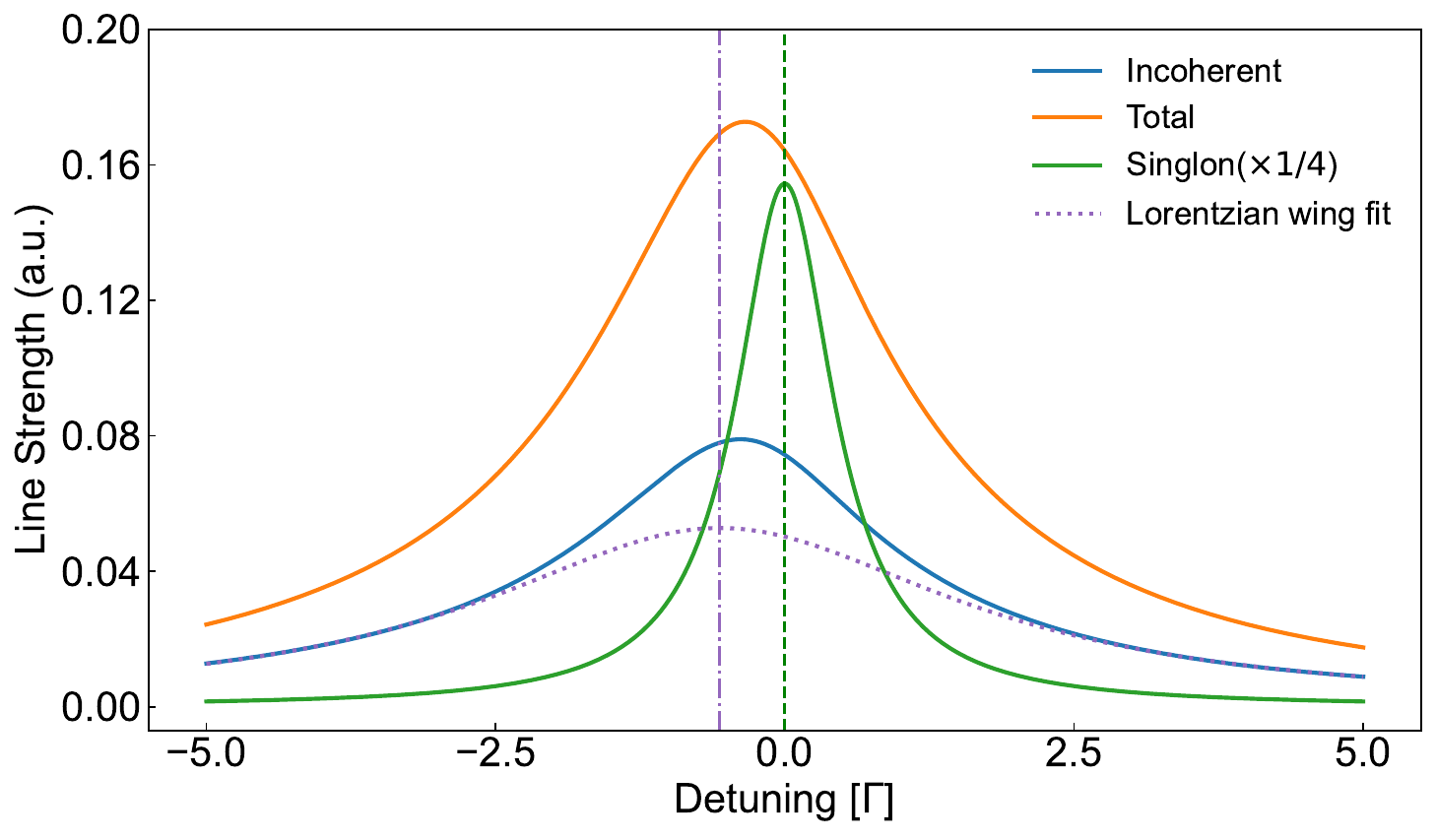}
    \caption{\textbf{Numerical simulation of fluorescence spectra}. The fluorescence spectrum is calculated for singlons and doublons with anisotropic confinement $(\sigma_\mathrm{h},\sigma_\mathrm{z})=(22 \,\text{nm},16 \,\text{nm})$. For doublons, the relative line strength is divided by 2, the atom number. The singlon spectrum is divided by a factor of 4 for display. When the dipole-dipole interaction is set to 0, the total fluorescence for doublons is twice larger than for singlons at large detuning due to the superradiance effect. The doublon curve is clearly broadened and shifted to the negative side in the tightly squeezed anisotropic Wannier function. The purple curve is a Lorentzian wing fit to the incoherent line strength curve for detuning $\Delta\in ([-5,-3]\cup[3,5]) \Gamma_{421}$.}
    \label{fig:lineshape}
\end{figure}

For the light scattered from the 3D atom array, most of the coherent light is suppressed by destructive interference. The light collected on the detector is mainly the incoherent component, whose intensity is given by $I_{\text{incoh}}=I_{\text{total}}-I_{\text{coh}}$. In the main text, we have plotted the incoherent fluorescence spectrum. Here in Fig.~\ref{fig:lineshape} we show the comparison between incoherent, total light scattering of doublons and total light scattering of singlons.

In the two-body Lamb shift experiment, we have an atomic array consisting of a singlon shell and doublon core. A simple model of the light scattering gives $I(\Delta)=f_{\mathrm{doub}}\left((S(q)D+1-D)I_{\mathrm{coh}}^{\mathrm{(doub)}}(\Delta)+I_{\mathrm{incoh}}^{\mathrm{(doub)}}(\Delta)\right)+f_{\mathrm{sing}}(S(q)D+1-D)I_{\mathrm{coh}}^{\mathrm{(sing)}}(\Delta)$. Here $S(q)\approx0.3$ is the structure factor of the atomic array, and $D\approx0.85$ is the Debye-Waller factor~\cite{zwhd-1k2t}. $f_{\mathrm{doub}}\approx f_{\mathrm{sing}}\approx0.5$ are atom number fractions of doublons and singlons. $I_{\mathrm{coh}}^{\mathrm{(doub)}}(\Delta)$, $I_{\mathrm{incoh}}^{\mathrm{(doub)}}(\Delta)$, and $I_{\mathrm{coh}}^{\mathrm{(sing)}}(\Delta)$ represent the normalized coherent fluorescence spectrum of doublons, incoherent fluorescence spectrum of doublons, and coherent fluorescence spectrum of singlons in Fig.~\ref{fig:lineshape}, respectively. This model reduces the relative Lamb shift in Fig.~\ref{fig:TwoAtom_Lamb}b from the two-atom model prediction $0.54\Gamma_{421}$ to $0.38\Gamma_{421}$.

We have evidence from light scattering and photoassociation that even at very weak probe laser power the Franck-Condon region is depleted. We have included this in a model where atoms within $\pm 2 \Gamma$ detuning from the Franck-Condon region have been excluded before calculating the fluorescence spectrum. This reduced
the relative line shift of the two-atom fluorescence spectrum only at the 10\% level.
\subsubsection{Lamb shift at higher lattice fillings}
Here we show that the intra-site Lamb shift for higher fillings $N$ scales approximately with $N-1$ and present experimental results where such larger Lamb shifts have been observed.
We describe $N$ two-level atoms at positions $r_i$ with $\left|r_i-r_j\right| \ll \lambdabar$, interacting with the dipole-dipole Hamiltonian $H=\sum_{i < j} H_{i j}$, with $H_{i j}=V_{\mathrm{dd}}\left(r_i-r_j\right)\left(\sigma_i^{+} \sigma_j^{-}+\text{h.c.}\right)$, with $\sigma^+_i=\ket{e_i}\bra{g_i}$.
Perturbative light scattering probes the single-excitation eigenstates $\left|e_i\right\rangle$ with energy $E_i=\left\langle e_i\right| H\left|e_i\right\rangle$. In principle, each of the eigenstates has a different resonance frequency.
However, if the separation between $E_i$ is smaller than the natural linewidth of the electronic excited states, the different resonances are not resolved, and the experiment measures the weighted average of the resonances $\langle E\rangle=\sum_i E_i P_i$. Here, the weights $P_i=\left|\left\langle e_i \mid \psi\right\rangle\right|^2$ are given by the overlap between the eigenstates $\ket{e_i}$ and the symmetric single-excited state $|\psi\rangle=(|e_1g_2 \cdots g_N\rangle+|g_1e_2 \cdots g_N\rangle+\cdots+|g_1\cdots g_{N-1} e_N\rangle)/\sqrt{N}$.
The weighted average of the resonances can be calculated as
\begin{equation}
\begin{aligned}
\langle E\rangle & =\sum_i\langle\psi| e_i \rangle\langle e_i|H| e_i \rangle\langle e_i|\psi\rangle \\
& =\sum_{i j}\langle\psi| e_i \rangle\langle e_i|H| e_j \rangle\langle e_j|\psi\rangle\\
& =\langle\psi| H|\psi\rangle.
\end{aligned}
\end{equation}
Here, we used $\left\langle e_i\right| H\left|e_j\right\rangle=E_i\delta_{i j}$ and the completeness relation $\mathbb{I}=\sum_i\left|e_i \rangle\langle e_i\right|$.
Finally, we arrive at $\langle E\rangle=\frac{2}{N} \sum_{i < j} V_\mathrm{dd}\left(r_i-r_j\right)$.

This result has to be averaged over the density distributions of the atoms.
If all atoms have the same density distribution, then the result is independent of $i$ and $j$: $\left\langle V_{\mathrm{dd}}\left(r_i-r_j\right)\right\rangle \equiv V_{\mathrm{dd}}$. The Lamb shift can now be expressed as $\langle E\rangle=\frac{2}{N} \times \frac{N(N-1)}{2} V_{\mathrm{dd}}=(N-1) V_{\mathrm{dd}}$. The Lamb shift is proportional to $N-1$, similar to a mean-field result for on-site interactions.

\begin{figure}
    \centering
    \includegraphics[width=\linewidth]{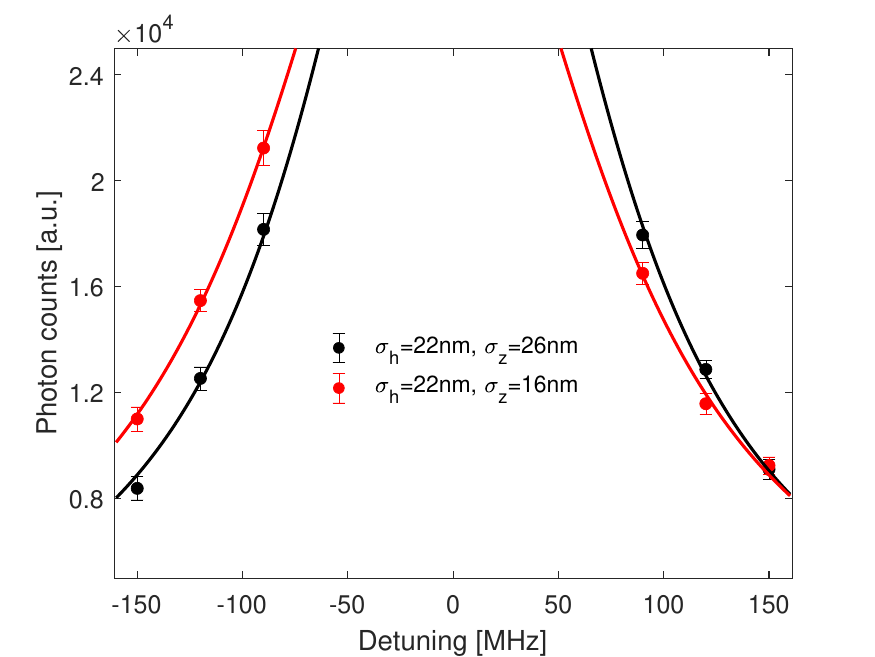}
    \caption{\textbf{Lamb shift at higher fillings.} We have measured the intrasite few-body Lamb shift at higher fillings (substantial triplon fraction), and obtained a relative shift of $12.7\pm1.8\mathrm{MHz}$ compared to $8.6\pm2.5\mathrm{MHz}$ for a mixture of singlons and doublons (see Fig.~\ref{fig:TwoAtom_Lamb}b). Each data point is averaged over independent experimental cycles, and the error bar reflects the s.e.m. }
    \label{fig:HigherFilling}
\end{figure}

Experimentally, we have measured the Lamb shift at higher lattice fillings by loading the optical lattices from a more tightly confined optical dipole trap. The observed shift is indeed larger (Fig.~\ref{fig:HigherFilling}).

\subsubsection{Calculation of the collective Lamb shift}

Since we use an array of atoms, there is the possibility of many-body Lamb shifts due to the $1/r$ retarded dipole-dipole interactions. Since this potential has $\exp(ikr)$ phase factors, the dominant contribution would come from groups of atoms along ``phase-matched'' directions, which are related to directions for Bragg scattering ~\cite{hutson2024observation,chang2004controlling}. The angle of the incident light is within 5 degrees from the closest Bragg angle, where a moderate collective shift on the order of $\Gamma_{421}$ occurs. Since the horizontal direction is closer to the Bragg condition, we kept the pair wavefunction constant along the horizontal direction and varied the vertical confinement. Using singlons we verified that this change of aspect ratio didn't cause any line shifts. Any collective Lamb shift would be accompanied by an angular pattern of light scattering caused by interference of light scattered by different atoms, which can be observed in time-of-flight as net momentum transfer to the atoms. We observed no changes when we varied the vertical confinement (but small changes for the horizontal case). These checks are strong evidence that the shift observed in Fig.~\ref{fig:TwoAtom_Lamb}b is due to the dipolar interactions within the pairs of atoms. For higher fillings in the optical lattice, we observed even larger Lamb shifts.

\begin{figure}
    \centering
    \includegraphics[width=1.0\linewidth]{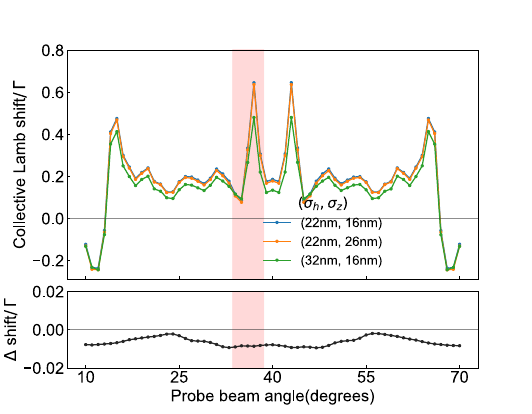}
    \caption{\textbf{Numerical simulations of the collective Lamb shift.} The probe beam incident angle is at $(36.0^\circ\pm 2.5^\circ)$, indicated by red shading.  At our probe beam angle, the maximal possible shift is about $0.6 (1.2) \Gamma_{421}$ for singlons (doublons). However, it varies between the blue and orange cloud sizes by less than $0.01 \Gamma$ (lower panel). For the same variation in cloud size for doublons, we observed a spectral shift of $0.27(8)\Gamma_{421}$ (see Fig.~\ref{fig:TwoAtom_Lamb}b) which is therefore interpreted as a two-body Lamb shift due to dipolar interactions between the atom pair. The curves are symmetric around 40$^\circ$, which is half the intersection angle of the horizontal lattice beams. } 
    \label{fig:Supp_CLS}
\end{figure}

We can validate these conclusions in a Monte Carlo simulation~\cite{hutson2024observation,chang2004controlling}. We calculate the collective Lamb shift in a $21\times21\times21$ lattice filled with singlons for our experimental lattice parameters with Gaussian position fluctuations which are different in different directions due to the anisotropy of the lattice. Fig.~\ref{fig:Supp_CLS} confirms that the collective Lamb shift is much more sensitive to the horizontal confinement, since the lattice is close to a Bragg condition along horizontal directions. Fig.~\ref{fig:Supp_CLS} also shows that for the differential Lamb shift measurements in Fig.~\ref{fig:TwoAtom_Lamb} and Fig.~\ref{fig:HigherFilling} the contribution of the collective Lamb shift (i.e., between lattice sites) accounts for less than 10 \% of the observed shift, which is therefore a two-body Lamb shift of the on-site atom pair.

\subsection{Simulation for photoassociative loss}

Here we describe how the experimental and two theoretical loss rate curves shown in Fig.\ref{fig:TwoAtom_PA} were obtained.
\subsubsection{Dynamical model}

We have performed semi-classical Monte-Carlo simulations for photoassociative using methods similar to \cite{PhysRevLett.134.013202}. In a simplified model, we assume an isotropic interaction potential between the excited state atom pair. Two classical $^{162}\mathrm{Dy}$ atoms are initialized in the harmonic-oscillator ground state on one site of the optical lattice. The pair of atoms is excited by the laser light, when they cross the FC region which brings the atoms into resonance. The scattering rate with a Lorentzian lineshape is integrated over the discrete time steps. If an atom is resonantly excited, the atom pair will experience the resonant dipole-dipole force for around one excited state lifetime. An atom is counted as lost when any one-dimensional oscillator energy, including both kinetic and potential energy, exceeds the corresponding finite lattice depth.  The simulation records the ensemble-averaged atom number
$\langle N(t)\rangle$ and converts its early-time decay into an effective
two-body coefficient $\beta$.

The initial positions and momenta for the atom pair $(\boldsymbol{r}_\mathrm{A},\boldsymbol{p}_\mathrm{A},\boldsymbol{r}_\mathrm{B},\boldsymbol{p}_\mathrm{B})$ are randomly chosen from the quantum harmonic oscillator ground state Wigner distribution. Initial samples with a total energy that already exceeds the finite lattice barrier are rejected. The simulation uses a laser saturation parameter $ s_{\rm PA}=2.3$ and a lattice trap depth of $(900 \,E_\mathrm{R,h},910\,E_\mathrm{R,h},200\,E_\mathrm{R,v})$, following the parameters in our experiment. Here, we take $\hat{y}-$lattice depth to be slightly different from $\hat{x}-$lattice to avoid closed classical elliptical trajectories. These trajectories have a certain minimal distance to the origin and would never cross the Franck-Condon region, which is unphysical for quantum wave packets. The detuning of the 421 nm light is varied from $\pm$1 GHz to $\pm$40 GHz.

For the simplicity of the simulation, the isotropic excited state atom pair $C_3$ potential is defined as
\begin{equation}
\label{eq:Viso}
    V_{\mathrm{iso}}(R\equiv|\boldsymbol{r}_A-\boldsymbol{r}_B|)= \pm \frac{3}{4}\frac{\hbar \Gamma_{421}}{k^3_{\mathrm{421}}}\frac{1}{R^3} \equiv \pm \frac{C_3}{R^3}.
\end{equation}
The positive(negative) signs stand for repulsive(attractive) $S+P$ branches. Since our photoassociation laser is detuned far away, we only consider the excitation to the repulsive(attractive) branch for blue(red) detuned laser. 

For an atom with a partner on the same lattice site, its instantaneous single-atom scattering rate is

\begin{equation}
W_{\rm sc}(R)
=
\frac{\Gamma_{421}}{2}
\frac{s_{\mathrm PA}}
{1+s_{\mathrm PA}
+
\left[2(\delta(R))/\Gamma_{421}\right]^2}
,
\label{eq:SWsc}
\end{equation}
which depends on the pair separation via $\delta(R)=\Delta-\mathrm{sgn}(\Delta)C_3/(\hbar R^3)$. We define the Condon radius by the condition that the dipolar shift compensates the laser detuning $R_C =({C_3}/{\hbar|\Delta|})^{1/3}.$
The simulation searches for crossings of the  Franck-Condon region within each geometry step $\Delta t$. In the simulation, $\Delta t$ is chosen to be around 0.3 $\mu s$, during which the atoms travel about 4 nm. This distance is much larger than the thickness of the Franck-Condon region, but smaller than the size of the atom density distribution. If a crossing event is detected, the relative coordinate is propagated under the harmonic trap to locate the first crossing time $\tau_C$. Near the crossing, the detuning is linearized as
\begin{equation}
\delta(t)
\simeq
\dot{\delta}(t-\tau_C),
\end{equation}
with
\begin{equation}
\dot{\delta}
=
3\,\operatorname{sgn}(\Delta)
\frac{C_3}{\hbar R_C^4}
\dot R,
\qquad
\dot R
=
\frac{\boldsymbol v_{\rm rel}\cdot \boldsymbol R}{R_C}.
\end{equation}

The integrated excitation rate of Eq. \ref{eq:SWsc} is  defined as $h$. For two atoms, it becomes $2h$. The probability for no excitation is $e^{-2h}$.
 Note that for detunings larger than 3 GHz, the Franck-Condon region is so thin that the classical transit time of the atoms becomes shorter than the excited-state lifetime. In this case, a Landau-Zener model provides a better description than the Fermi's Golden Rule expression in Eq. \ref{eq:SWsc}. For small excitation rates, both models give exactly the same total excitation rate. We have used Eq. \ref{eq:SWsc} for all parameters in Fig. \ref{fig:TwoAtom_PA}, somewhat overestimating the role of saturation by at most $\sqrt{1+s_{\rm PA}}\approx 1.8$.

At the time of absorption during a crossing event, the atom pair is transferred to the excited state and remains there for about $5\, {\rm ns}$.
The excited pair evolves under the lattice force and the 
resonant dipole-dipole force $\boldsymbol {F}_{\rm dd}= -\nabla V_{\rm iso}$ which provides a large momentum kick leading to escape of the atom pair. 

In the simulation, if no crossing event is detected, the atoms may still receive momentum kicks at random times due to absorption and emission events with a much lower rate by taking $V_{\rm iso}=0$ in Eq. \ref{eq:Viso}. The recoil of the absorbed photons is along the $\hat{z}$ direction, while the emission direction is randomly distributed over the full solid angle. These momentum kicks are important to randomize the orbits for a spherical trap, but don't have a significant effect on the loss rate for anisotropic traps (increase by up to 50\%).

The recorded observable is the mean number of trapped atoms, $\langle N(t)\rangle$. Because each simulation trajectory contains only one atom pair, the decay is fit as an effective single-pair survival process rather than to the continuum two-body form proportional to $N^2$. We therefore use
\begin{equation}
\langle N(t)\rangle
=
N_0 e^{-\gamma_{\rm pair} t}.
\end{equation}
The corresponding effective two-body loss coefficient is reported as
\begin{equation}
\beta
=
\gamma_{\rm pair}{V_{\rm eff}},
\end{equation}
where $V_{\rm eff}=(8\pi^{3/2}\sigma_x\sigma_y\sigma_z)$ is the effective volume determined by the ground state widths of the harmonic oscillator. The fit is performed over the decay to 50\% population. 

\subsubsection{Static GP model}

In a seminal paper, Gallagher and Pritchard (GP) formulated a model for photo-associative loss~\cite{Gallagher1989} which is used here as a benchmark comparison to our experimental data. We consider the same simplified two-level system, where the excited state experiences an isotropic molecular potential $V_{\rm iso}(R)$. The atom pair is considered static with a uniform density distribution in the relative coordinate $\{\boldsymbol{R}\}$. The atom loss can be described by two events in rapid succession: photo-association of the atoms into an excited molecular state, and the atom pair loss after de-excitation. 

\begin{equation}
    \beta
=
4\pi
\int_0^\infty
dR\,
R^2
\Gamma_{\mathrm{asso}}(R)P_{\mathrm{loss}}(R).
\end{equation}

The molecule association process rate is given by the steady-state scattering rate with a Lorentzian linewidth
$\Gamma_{\mathrm{asso}}(R)=2W_{sc}(R)$.
The loss rate is obtained by evolving the excited state pair on the molecular potential branch for around one excited state lifetime~\cite{Gallagher1989,dreon2017optical}. The result of the integral is shown in Fig.\ref{fig:TwoAtom_PA}. This model overestimates the loss rate by a factor between 5 and 40, which we interpret as being due to loss saturation of the incoming flux and dynamic depletion of the Franck-Condon region (see discussion later).

\subsubsection{Experimental data fit}
The measured two-body loss rates at different 421 nm detunings and laser intensities in Fig.\ref{fig:TwoAtom_PA} are obtained from fits to atom number decay curves, including the expansion of the atom cloud by the observed heating possibly induced by single photon recoil kicks. We illuminate the atom pairs and measure the remaining atom number and momentum width after a $T_{\rm tof}=3$ ms time-of-flight. The cloud-size evolution was obtained from the measured widths. For both image-plane directions, the widths are fitted to $w_j(t) =\sqrt{w_{j,0}^2+\alpha_j t},\,j\in\{x,z\}$. The fitted widths were converted to in-trap sizes $\sigma_x$ and $\sigma_z$ according to harmonic oscillator relation $\sigma(t) \sim w(t)/(\omega T_{\rm tof})$. The effective volume entering the loss equation was then $V(t)=2\sqrt{2}\,\sigma_x^2(t)\sigma_z(t)$. To get normalized atom number per doublon lattice site, we subtract the atom number contribution from singlons and normalize the initial atom number to be exactly two.

The normalized atom number is fitted with a exponential loss model
\begin{equation}
    \frac{dN}{dt}
    =
    -
    \frac{\beta}{(2\pi)^{3/2}V(t)}N ,
\end{equation}
where \(\beta\) is the fitted loss coefficient.  The differential equation was
 numerically integrated for each trial value of the fit parameters. For most of the data, the temperature of the cloud increased by less than a factor of two.
 
\subsubsection{Interpretation}

 The dynamical model simulations give loss rates close to the observed ones, whereas the static GP model predictions are around 30 times higher. The GP model is equivalent to a Landau-Zener crossing model for small  Landau-Zener parameters $\Lambda=\pi\hbar\Omega^2/|v_{\mathrm{rel}}U'(R_c)|\ll 1$~\cite{RevModPhys.71.1}. Large values of $\Lambda$ would suppress losses to below the predictions of the GP model.  However, our experimental values for $\Lambda$ are smaller than 1, except for the smallest detunings, and therefore Landau-Zener saturation can explain only a small part of the observed suppression of photoassociation loss rates.

Light scattering experiments show the suppression of inelastic light scattering originating from the FC region (see Fig.~\ref{fig:TwoAtom_LIDD}c for short laser pulses). The same mechanism should also suppress photoassociation. Indeed, light scattering experiments show the same detuning asymmetry as in Fig.~\ref{fig:TwoAtom_PA} for photoassociation loss. Other evidence for depletion comes from the shape of loss curves during photoassociation: several traces show a slowdown after a sizeable fraction of the atoms have been lost.

This interpretation is supported by our dynamical simulation of atom trajectories which show significant phase space depletion. For long illumination times, the simulated density distributions are no longer Gaussian, even for Landau-Zener parameters $\Lambda\ll1$. This can be understood by considering atoms with large, nearly circular orbits in three-dimensional space. These pairs circulate around the small Franck-Condon region and cross it much less often, with an effective crossing frequency on the order of the difference between the trap frequencies along $\hat{x},\hat{y},\hat{z}$. Thus, the later population is strongly survivor-selected and decreases slower. The simulated loss curve also deviates from the ideal exponential loss curve. Note that non-exponential decay has been observed by photoassociation of two atoms in tweezers \cite{PhysRevLett.126.083401}. However, in their case, it is due to a statistical ensemble of nodal and peaked states. In our case, we always start with symmetric peaked states, and we are not on resonance with a particular free atoms to molecular bound state transition.

The light-induced dipole-dipole interaction in Eq. \ref{eq:S7} creates a repulsive conservative force. For our experimental parameters, this potential barrier has a height of tens of MHz.  In a classical simulation, it would completely suppress photoassociation.  However, this potential barrier is very narrow (less than 0.1 nm for a laser detuning of 10 GHz), and therefore the tunneling transmission is close to one. We have therefore performed the semiclassical simulations without the light induced dipole-dipole force.

One would expect that at much lower laser intensity depletion no longer occurs. In fact, at the lowest loss rates, some decay curves showed a simple exponential form, and for the lowest laser intensities more inelastic light scattering was observed (see Fig.~\ref{fig:TwoAtom_LIDD}c).  It seems that the wavefunction dynamics is rich and sensitively depends on detuning and laser intensity in a complex way. Since these effects have not been observed before, they are probably related to the new regime of high densities realized in our experiment, where the rate of excitation in the thin Franck-Condon region dominates over the excitation rate integrated over the rest of the cloud.

\end{document}